\documentclass[12pt]{article}
\usepackage{myart}

\renewcommand{\theequation}{\arabic{section}.\arabic{equation}}
\normalbaselines
\input tcilatex.tex

\begin{document}

\author{Yuri A. Rylov}
\title{Is the Dirac particle composite?}
\date{Institute for Problems in Mechanics, Russian Academy of Sciences \\
101-1 ,Vernadskii Ave., Moscow, 119526, Russia \\
email: rylov@ipmnet.ru\\
Web site: {$http://rsfq1.physics.sunysb.edu/\symbol{126}rylov/yrylov.htm$}\\
or mirror Web site: {$http://195.208.200.111/\symbol{126}rylov/yrylov.htm$}}
\maketitle

\begin{abstract}
Classical model $\mathcal{S}_{\mathrm{Dcl}}$ of the Dirac particle $\mathcal{%
S}_{\mathrm{D}}$ is constructed. $\mathcal{S}_{\mathrm{D}}$ is the dynamic
system described by the Dirac equation. For investigation of $\mathcal{S}_{%
\mathrm{D}}$ and construction of $\mathcal{S}_{\mathrm{Dcl}}$ one uses a new
dynamic method: dynamic disquantization. This relativistic purely dynamic
procedure does not use principles of quantum mechanics. The obtained
classical analog $\mathcal{S}_{\mathrm{Dcl}}$ is described by a system of
ordinary differential equations, containing the quantum constant $\hbar $ as
a parameter. Dynamic equations for $\mathcal{S}_{\mathrm{Dcl}}$ are
determined by the Dirac equation uniquely. The dynamic system $\mathcal{S}_{%
\mathrm{Dcl}}$ has ten degrees of freedom and cannot be a pointlike
particle, because it has an internal structure. There are two ways of
interpretation of the dynamic system $\mathcal{S}_{\mathrm{Dcl}}$: (1)
dynamical interpretation and (2) geometrical interpretation. In the
dynamical interpretation the classical Dirac particle $\mathcal{S}_{\mathrm{%
Dcl}}$ is a two-particle structure (special case of a relativistic rotator).
It explains freely such properties of $\mathcal{S}_{\mathrm{D}}$ as spin and
magnetic moment, which are strange for pointlike structure. In the
geometrical interpretation the world tube of $\mathcal{S}_{\mathrm{Dcl}}$ is
a ''two-dimensional broken band'', consisting of similar segments. These
segments are parallelograms (or triangles), but not the straight line
segments as in the case of a structureless particle. Geometrical
interpretation of the classical Dirac particle $\mathcal{S}_{\mathrm{Dcl}}$
generates a new approach to the elementary particle theory.
\end{abstract}


\newpage

\section{Introduction}

The Dirac particle is the dynamic system $\mathcal{S}_{\mathrm{D}}$,
described by the Dirac equation. This is one of wide-spread dynamic systems
used in theory of quantum phenomena. Mathematical analysis of properties of
the dynamic system $\mathcal{S}_{\mathrm{D}}$ is of undoubted interest. The
Dirac dynamic system $\mathcal{S}_{\mathrm{D}}$ was investigated by many
researchers. There is no possibility to list all them, and we mention only
some of them. First, this is transformation of the Dirac equation on the
base of quantum mechanics \cite{D58,FW50}. The complicated structure of
Dirac particle was discovered by Schr\"{o}dinger \cite{S930}, who
interpreted it as some complicated quantum motion (zitterbewegung).
Investigation of this quantum motion and different models of Dirac particle
can be found in \cite{B84,BB81,A81,H90,RV93} and references therein. Our
investigation differs in absence of any suppositions on the Dirac particle
model and in absence of referring to the quantum principles. We use only
dynamic methods and \textit{investigate the Dirac particle simply as a
dynamic system.}

Conventionally the analysis of the dynamic system $\mathcal{S}_{\mathrm{D}}$
and its dynamic equations is carried out by a use of the quantum mechanics
principles. In particular, it means that the quantum constant $\hbar $ is
not simply a parameter of the dynamic system $\mathcal{S}_{\mathrm{D}}$. The
quantum constant $\hbar $ is provided by some additional physical meaning.
It is supposed that, if $\hbar \rightarrow 0$, all quantum effects are cut
off, and dynamic system $\mathcal{S}_{\mathrm{D}}$ turns into classical
dynamic system $\mathcal{S}_{\mathrm{Dcl}}$, having six degrees of freedom.

Usually it is supposed that $\mathcal{S}_{\mathrm{Dcl}}$ is a pointlike
relativistic particle of mass $m$, having a spin (angular momentum) $S_{%
\mathrm{D}}=\hbar /2$ and magnetic momentum $\mu _{\mathrm{D}}=e\hbar /mc$.
Procedure of transition from $\mathcal{S}_{\mathrm{D}}$ to $\mathcal{S}_{%
\mathrm{Dcl}}$ is called the transition to classical description. However,
the transition to the limit $\hbar \rightarrow 0$ is not carried out in such
quantities as spin $S_{\mathrm{D}}=\hbar /2$ and magnetic momentum $\mu _{%
\mathrm{D}}=e\hbar /mc$, which remain to be quantum in the sense that they
contain the unvanishing quantum constant $\hbar $. One states that spin and
magnetic moment are quantities, which have no classical analog.

In addition a direct transition to the limit $\hbar \rightarrow 0$ in the
action for the dynamic system $\mathcal{S}_{\mathrm{D}}$ is impossible.
Indeed, the action $\mathcal{A}_{\mathrm{D}}$ for the dynamic system $%
\mathcal{S}_{\mathrm{D}}$ has the form 
\begin{equation}
\mathcal{S}_{\mathrm{D}}:\qquad \mathcal{A}_{\mathrm{D}}[\bar{\psi},\psi
]=\int (-m\bar{\psi}\psi +\frac{i}{2}\hbar \bar{\psi}\gamma ^{l}\partial
_{l}\psi -\frac{i}{2}\hbar \partial _{l}\bar{\psi}\gamma ^{l}\psi )d^{4}x
\label{b1.1}
\end{equation}
Here $\psi $ is four-component complex wave function, $\psi ^{\ast }$ is the
Hermitian conjugate wave function, and $\bar{\psi}=\psi ^{\ast }\gamma ^{0}$
is conjugate one. $\gamma ^{i}$, $i=0,1,2,3$ are $4\times 4$ complex
constant matrices, satisfying the relation 
\begin{equation}
\gamma ^{l}\gamma ^{k}+\gamma ^{k}\gamma ^{l}=2g^{kl}I,\qquad k,l=0,1,2,3.
\label{b1.2}
\end{equation}
where $I$ is the unit $4\times 4$ matrix, and $g^{kl}=$diag$\left(
c^{-2},-1,-1,-1\right) $ is the metric tensor. Considering dynamic system $%
\mathcal{S}_{\mathrm{D}}$, we choose for simplicity such units, where the
speed of the light $c=1$. The action (\ref{b1.1}) generates dynamic equation
for the dynamic system $\mathcal{S}_{\mathrm{D}}$, known as the Dirac
equation 
\begin{equation}
i\hbar \gamma ^{l}\partial _{l}\psi -m\psi =0  \label{f1.2}
\end{equation}
and expressions for physical quantities: the 4-flux $j^{k}$ of particles and
the energy-momentum tensor $T_{l}^{k}$%
\begin{equation}
j^{k}=\bar{\psi}\gamma ^{k}\psi ,\qquad T_{l}^{k}=\frac{i}{2}\left( \bar{\psi%
}\gamma ^{k}\partial _{l}\psi -\partial _{l}\bar{\psi}\cdot \gamma ^{k}\psi
\right)  \label{f1.3}
\end{equation}
If we set $\hbar =0$ in the action (\ref{b1.1}), we obtain no description.
For transition to the classical description, where $\hbar =0$, we need more
subtle methods.

For simplicity we consider a use of these methods in the simple example of
the dynamic system $\mathcal{S}_{\mathrm{S}}$, described by the
Schr\"{o}dinger equation. In this case the action has the form 
\begin{equation}
\mathcal{S}_{\mathrm{S}}:\qquad \mathcal{A}_{\mathrm{S}}\left[ \psi ,\psi
^{\ast }\right] =\int \left\{ \frac{i\hbar }{2}\left( \psi ^{\ast }\partial
_{0}\psi -\partial _{0}\psi ^{\ast }\cdot \psi \right) -\frac{\hbar ^{2}}{2m}%
\mathbf{\nabla }\psi ^{\ast }\mathbf{\nabla }\psi \right\} dtd\mathbf{x}
\label{b1.3}
\end{equation}

Expressions of the 4-current $j^{k}$ and components $T_{k}^{0}$ of
energy-momentum tensor have the form 
\begin{equation}
j^{k}=\left\{ \rho ,\mathbf{j}\right\} ,\qquad \rho =\psi ^{\ast }\psi
,\qquad \mathbf{j}=-\frac{i\hbar }{2m}\left( \psi ^{\ast }\mathbf{\nabla }%
\psi -\mathbf{\nabla }\psi ^{\ast }\cdot \psi \right)  \label{b1.2a}
\end{equation}
\begin{equation}
T_{0}^{0}=\frac{\hbar ^{2}}{2m}\mathbf{\nabla }\psi ^{\ast }\mathbf{\nabla }%
\psi ,\qquad T_{\alpha }^{0}=-\frac{i\hbar }{2}\left( \psi ^{\ast }\partial
_{\alpha }\psi -\partial _{\alpha }\psi ^{\ast }\cdot \psi \right) ,\qquad
\alpha =1,2,3  \label{b1.2b}
\end{equation}

If we set $\hbar =0$ in the action (\ref{b1.3}), we do not obtain classical
description of anything. To obtain the true result, we are to make at first
the transformation of the wave function phase 
\begin{equation}
\Psi =\exp \left( \frac{\hbar }{b_{0}}\ln \frac{\psi }{\left| \psi \right| }%
\right) \left| \psi \right| ,\qquad \psi =\exp \left( \frac{b_{0}}{\hbar }%
\ln \frac{\Psi }{\left| \Psi \right| }\right) \left| \Psi \right|
\label{b1.4}
\end{equation}
in the action (\ref{b1.3}). Here $b_{0}\neq 0$ is an arbitrary real
constant. After transformation the action (\ref{b1.3}) takes the form 
\[
\mathcal{S}_{\mathrm{S}}:\qquad \mathcal{A}_{\mathrm{S}}\left[ \Psi ,\Psi
^{\ast }\right] =\int \left\{ \frac{ib_{0}}{2}\left( \Psi ^{\ast }\partial
_{0}\Psi -\partial _{0}\Psi ^{\ast }\cdot \Psi \right) -\frac{b_{0}^{2}}{2m}%
\mathbf{\nabla }\Psi ^{\ast }\mathbf{\nabla }\Psi \right. 
\]
\begin{equation}
-\left. \frac{\hbar ^{2}-b_{0}^{2}}{8\Psi ^{\ast }\Psi }\left( \mathbf{%
\nabla }\left( \Psi ^{\ast }\Psi \right) \right) ^{2}\right\} dtd\mathbf{x}
\label{b1.5}
\end{equation}

The 4-current (\ref{b1.2a}) and components of the energy momentum tensor (%
\ref{b1.2b}) take the form 
\begin{equation}
j^{k}=\left\{ \rho ,\mathbf{j}\right\} ,\qquad \rho =\Psi ^{\ast }\Psi
,\qquad \mathbf{j}=-\frac{ib_{0}}{2m}\left( \Psi ^{\ast }\mathbf{\nabla }%
\Psi -\mathbf{\nabla }\Psi ^{\ast }\cdot \Psi \right)  \label{b1.5a}
\end{equation}
\begin{equation}
T_{0}^{0}=\frac{b_{0}^{2}}{2m}\mathbf{\nabla }\Psi ^{\ast }\mathbf{\nabla }%
\Psi +\frac{\hbar ^{2}-b_{0}^{2}}{8\Psi ^{\ast }\Psi }\left( \mathbf{\nabla }%
\left( \Psi ^{\ast }\Psi \right) \right) ^{2},  \label{b1.5b}
\end{equation}
\begin{equation}
T_{\alpha }^{0}=-\frac{ib_{0}}{2}\left( \Psi ^{\ast }\partial _{\alpha }\Psi
-\partial _{\alpha }\Psi ^{\ast }\cdot \Psi \right) ,\qquad \alpha =1,2,3
\label{b1.5c}
\end{equation}
If now we set $\hbar =0$ in the action (\ref{b1.5}), we obtain the action
for the pure statistical ensemble $\mathcal{E}\left[ \mathcal{S}_{\mathrm{Scl%
}}\right] $ of dynamic systems $\mathcal{S}_{\mathrm{Scl}}$. The action for
the dynamic system $\mathcal{S}_{\mathrm{Scl}}$ has the form$\mathcal{\ }$%
\begin{equation}
\mathcal{S}_{\mathrm{Scl}}:\qquad \mathcal{A}_{\mathrm{Scl}}\left[ \mathbf{x}%
\right] =\int \frac{m}{2}\left( \frac{d\mathbf{x}}{dt}\right) ^{2}dt
\label{b1.6}
\end{equation}
where $\mathbf{x}=\mathbf{x}\left( t\right) =\left\{ x^{1}\left( t\right)
,x^{2}\left( t\right) ,x^{3}\left( t\right) \right\} $. The action (\ref
{b1.6}) describes the free classical nonrelativistic particle.

Thus, the continuous dynamic system $\mathcal{S}_{\mathrm{S}}$, having
infinite number of the freedom degrees, associates with the discrete dynamic
system $\mathcal{S}_{\mathrm{Scl}}$, having six degrees of freedom. This
circumstance is formulated as follows. Dynamic system $\mathcal{S}_{\mathrm{S%
}}$ is a result of quantization of the free nonrelativistic particle $%
\mathcal{S}_{\mathrm{Scl}}$. One may say also, that the classical dynamic
system $\mathcal{S}_{\mathrm{Scl}}$ is a result of disquantization of the
quantum system $\mathcal{S}_{\mathrm{S}}$.

We note two important properties of the disquantization, i.e. transition
from the continuous dynamic system $\mathcal{S}_{\mathrm{S}}$ to the
discrete dynamic system $\mathcal{S}_{\mathrm{Scl}}$.

1. The intermediate dynamic system (\ref{b1.5}) is not quantum, because the
dynamic equation for the wave function $\Psi $ is nonlinear (but this
disagrees with the quantum mechanics principles). Nonlinearity of dynamic
equation for $\Psi $ is connected with the fact that transformation (\ref
{b1.4}), connecting the wave function $\psi $ and $\Psi $ is nonlinear. Of
course, the accordance with the quantum principles must be violated at some
moment, as far as the final dynamic system $\mathcal{S}_{\mathrm{Scl}}$ is
not quantum. But it is curious that the accordance is violated not in the
time, when we set $\hbar =0$, but at the earlier stage, when the action (\ref
{b1.3}) is transformed to the action (\ref{b1.5}), describing the same
dynamic system $\mathcal{S}_{\mathrm{S}}$, as the action (\ref{b1.3}).

2. The transformation (\ref{b1.4}), connecting wave functions $\Psi $ and $%
\psi $ contains the quantum constant $\hbar $. It becomes to be singular at $%
\hbar \rightarrow 0$. It means essentially, that the quantum constant $\hbar 
$ is introduced in dynamic variables $\Psi $ and $\Psi ^{\ast }$.
Hereinafter, when $\hbar \rightarrow 0$, the quantum constant, contained in $%
\Psi $ is not changed. In a similar manner we act at disquantization of the
dynamic system $\mathcal{S}_{\mathrm{D}}$, when we introduce the quantum
constant in definition of spin $S_{\mathrm{D}}=\hbar /2$ and magnetic moment 
$\mu _{\mathrm{D}}=e\hbar /mc$. These quantities are not changed, when we go
to the limit $\hbar \rightarrow 0$.

Finally, why do we choose the transformation (\ref{b1.4}) for
disquantization of the dynamic system $\mathcal{S}_{\mathrm{S}}$, but not
some other? What motives are used at the choice of the transformation (\ref
{b1.4})? What transformation should be chosen for disquantization of the
dynamic system $\mathcal{S}_{\mathrm{D}}$?

We know that the dynamic system (\ref{b1.5}) is a result of quantization of
the dynamic system (\ref{b1.6}). The disquantization is the operation
reciprocal to the quantization. So we choose the manner of introduction of
the quantum constant $\hbar $ into the dynamic variable (wave function) in
such a way, to obtain (\ref{b1.6}) as a result of disquantization of dynamic
system $\mathcal{S}_{\mathrm{S}}$, described by the action (\ref{b1.3}).

In the case of the Dirac equation the situation is another one. The dynamic
system $\mathcal{S}_{\mathrm{D}}$ was postulated by Dirac. It was not
obtained as a result of quantization of some classical dynamic system. The
fact that the result of disquantization of $\mathcal{S}_{\mathrm{D}}$ is a
relativistic particle with a spin (i.e. the relativistic generalization of
the nonrelativistic particle with a spin, described by the Pauli equation)
was considered to be evident. One needs only to invent the proper
disquantization procedure.

At the disquantization of the dynamic system $\mathcal{S}_{\mathrm{D}}$
different authors use different methods \cite{P32,RK63,BK99,S2000}. All this
forces one to think that there is no general principle of the dynamic system
disquantization. The classical dynamic system (a result of disquantization
of quantum system) is obtained from some a priori consideration, and
corresponding methods of disquantization are fitted to the a priori result.
The common feature of all methods of disquantization is the fact, that the
dynamic equations for the disquantized dynamic system do not contain quantum
constant $\hbar $, although physical quantities (spin, magnetic moment) may
contain unvanishing quantum constant. It is common practice to think that
the dynamic system $\mathcal{S}_{\mathrm{Dcl}}$, obtained as a result of
disquantization of $\mathcal{S}_{\mathrm{D}}$ is a pointlike relativistic
particle with the spin $S_{\mathrm{D}}=\hbar /2$ and magnetic moment $\mu _{%
\mathrm{D}}=e\hbar /mc$. This belief has historical origin, and one cannot
substantiate it mathematically, because the result depends on the applied
methods.

Do there exist the principle of disquantization, which satisfies the
following conditions?

1. The quantum principles and a reference to the quantum constant are not
used. In particular, the transition to the limit $\hbar \rightarrow 0$ and
the introduction of the quantum constant into dynamic variables, connected
with this transition, is not used.

2. Disquantization of continuous dynamic system $\mathcal{S}$ determines
uniquely a disquantized classical dynamic system $\mathcal{S}_{\mathrm{cl.}}$%
.

If one succeeded to define the disquantization procedure in accordance with
these conditions, this procedure will be the means of investigation of
continuous dynamic system $\mathcal{S}$ and will realize an interpretation
of $\mathcal{S}$ in terms of the discrete classical system $\mathcal{S}_{%
\mathrm{cl}}$.

To solve this problem we need at first to perceive that the dynamic system $%
\mathcal{S}_{\mathrm{cl}}$ is classical and suitable for interpretation of $%
\mathcal{S}$, only in the case, when $\mathcal{S}_{\mathrm{cl}}$ has finite
number of the freedom degrees and, hence, its dynamic equations are ordinary
differential equations. It is of no importance whether or not these dynamic
equations contain the quantum constant $\hbar $, because effectiveness and
simplicity of the dynamic equations analysis is connected with the fact that
these equations are ordinary differential equations, but not partial
differential equations.

For the system of partial differential equations to be equivalent to a
system of ordinary differential equations, it is necessary that the
equations contain derivative only in one direction in the space of
independent variables. This direction is the direction of the current
4-vector $j^{k}$, which is determined by the dynamic system $\mathcal{S}$.
We choose dependent variables of dynamic system $\mathcal{S}$ in such a way,
that variables $j^{k}$, $k=0,1,2,3$ are among them, and make in dynamic
equations the change 
\begin{equation}
\partial ^{l}\rightarrow \partial _{||}^{l}=\frac{j^{l}j^{k}}{j^{s}j_{s}}%
\partial _{k},\qquad l=0,1,2,3,\qquad \partial ^{k}\equiv g^{kl}\partial
_{l}\equiv g^{kl}\frac{\partial }{\partial x_{l}},\qquad j_{l}\equiv
g_{lk}j^{k}  \label{b1.8}
\end{equation}
where $g_{ik}=$diag$\left\{ c^{2},-1,-1,-1\right\} $, $g^{ik}=\left\{
c^{-2},-1,-1,-1\right\} $. The dynamic equations of the system $\mathcal{S}$
turn into a system of ordinary differential equations. It is the system of
dynamic equation for the pure statistical ensemble $\mathcal{E}\left[ 
\mathcal{S}_{\mathrm{cl}}\right] $. The dynamic system $\mathcal{S}_{\mathrm{%
cl}}$ is classical in the sense, that it has a finite number of the freedom
degrees, i.e. its dynamic equations are ordinary differential equations.

Let us make a change of variables in the action (\ref{b1.5}) 
\begin{equation}
\Psi =\sqrt{\rho }e^{i\varphi },\qquad \Psi ^{\ast }=\sqrt{\rho }%
e^{-i\varphi }  \label{b1.7}
\end{equation}
We obtain instead of (\ref{b1.5}) 
\begin{equation}
\mathcal{S}_{\mathrm{S}}:\qquad \mathcal{A}_{\mathrm{S}}\left[ \rho ,\varphi %
\right] =\int \left\{ -b_{0}\left( j^{0}\partial _{0}\varphi +\mathbf{%
j\nabla }\varphi \right) -\frac{\hbar ^{2}}{8m}\frac{\left( \mathbf{\nabla }%
\rho \right) ^{2}}{\rho }\right\} dtd\mathbf{x}  \label{b1.8a}
\end{equation}
where in accordance with (\ref{b1.7}) and (\ref{b1.5a}) 
\begin{equation}
j^{0}=\rho ,\qquad \mathbf{j=}\frac{b_{0}}{2m}\rho \mathbf{\nabla }\varphi
\label{b1.9}
\end{equation}

We make the change (\ref{b1.8}) in (\ref{b1.8a}). The first term have the
form $j^{k}\partial _{k}\varphi $, and the procedure (\ref{b1.8}) does not
change it. For the second term we have 
\begin{equation}
\frac{\left( \mathbf{\nabla }\rho \right) ^{2}}{\rho }\rightarrow \frac{%
\mathbf{j}^{2}\left( j^{k}\partial _{k}\rho \right) }{\rho \left(
j^{s}j_{s}\right) ^{2}}=\frac{\left( \frac{b_{0}}{2m}\right) ^{2}\rho
^{2}\left( \mathbf{\nabla }\varphi \right) ^{2}\left( \rho \partial _{0}\rho
+\frac{b_{0}}{2m}\rho \left( \mathbf{\nabla }\varphi \right) \mathbf{\nabla }%
\rho \right) ^{2}}{\rho \left( c^{2}\rho ^{2}-\left( \frac{b_{0}}{2m}\right)
^{2}\rho ^{2}\left( \mathbf{\nabla }\varphi \right) ^{2}\right) ^{2}}%
=O\left( c^{-4}\right)  \label{b1.10}
\end{equation}

In the nonrelativistic approximation, when $c\rightarrow \infty $, this term
vanishes. It vanishes independently of whether or not $\hbar \rightarrow 0$.
But in the relativistic case the result may be different, and dependence of
the action on the quantum constant $\hbar $ may remain. We shall see that in
the case of the dynamic system $\mathcal{S}_{\mathrm{D}}$ the quantum
constant $\hbar $ remains in the action after the procedure (\ref{b1.8}).
Nevertheless the obtained dynamic system is a pure statistical ensemble of
classical dynamic systems $\mathcal{S}_{\mathrm{Dcl}}$, whose dynamic
equations are ordinary differential equations.

In the nonrelativistic approximation the action (\ref{b1.8a}) takes the form 
\begin{equation}
\mathcal{E}\left[ \mathcal{S}_{\mathrm{Scl}}\right] :\qquad \mathcal{A}_{%
\mathrm{Scl}}\left[ \rho ,\varphi \right] =\int \left\{ -b_{0}\rho \partial
_{0}\varphi -\frac{b_{0}^{2}}{2m}\rho \left( \mathbf{\nabla }\varphi \right)
^{2}\right\} dtd\mathbf{x}  \label{b1.11}
\end{equation}
The action (\ref{b1.11}) generates dynamic equations 
\begin{equation}
\frac{\delta \mathcal{A}}{\delta \varphi }=b_{0}\left( \partial _{0}\rho +%
\mathbf{\nabla }\left( \frac{\rho }{m}\mathbf{\nabla }\left( b_{0}\varphi
\right) \right) \right) =b_{0}\left( \partial _{0}j^{0}+\mathbf{\nabla j}%
\right) =0  \label{b1.12}
\end{equation}
\begin{equation}
\frac{\delta \mathcal{A}}{\delta \rho }=\partial _{0}\left( b_{0}\rho
\right) +\frac{1}{2m}\left( \mathbf{\nabla }\left( b_{0}\varphi \right)
\right) ^{2}=0  \label{b1.13}
\end{equation}

Equation (\ref{b1.12}) is the continuity equation, and (\ref{b1.13}) is the
Hamilton-Jacobi equation for the free nonrelativistic particle, where $%
b_{0}\varphi $ is the action variable. Introducing Lagrangian variables, one
can show \cite{R2004} that the action (\ref{b1.11}) is a special
(irrotational) case of the action 
\begin{equation}
\mathcal{E}\left[ \mathcal{S}_{\mathrm{Scl}}\right] :\qquad \mathcal{A}_{%
\mathcal{E}\left[ \mathcal{S}_{\mathrm{Scl}}\right] }\left[ \mathbf{x}\right]
=\int \frac{m}{2}\left( \frac{d\mathbf{x}}{dt}\right) ^{2}dtd\mathbf{\xi }
\label{b1.13a}
\end{equation}
where $\mathbf{x}=\mathbf{x}\left( t,\mathbf{\xi }\right) $, and $\mathbf{%
\xi }=\left\{ \xi _{1},\xi _{2},\xi _{3}\right\} $ are Lagrangian
coordinates, labelling the systems $\mathcal{S}_{\mathrm{Scl}}$,
constituting the statistical ensemble. Formally dynamic equations for the
statistical ensemble $\mathcal{E}\left[ \mathcal{S}_{\mathrm{Scl}}\right] $
are partial differential equations with independent variables $t,\xi
_{1},\xi _{2},\xi _{3}$. But in fact they contain only derivatives with
respect to the variable $t$, and dynamic equations are ordinary differential
equations, because independent variables $\mathbf{\xi }$ are contained in
dynamic equations as parameters (in reality dynamic equations do not depend
on $\mathbf{\xi }$ explicitly). Formal transformation of the action (\ref
{b1.11}) to the form (\ref{b1.13a}) is not simple, because the irrotational
flow (the property of the dynamic system (\ref{b1.11})) is described rather
easily in the Eulerian coordinates $\left\{ t,\mathbf{x}\right\} $, but its
expression in the Lagrangian coordinates $\left\{ t,\mathbf{\xi }\right\} $
is not simple.

If we apply transformation 
\begin{equation}
\psi =\sqrt{\rho }e^{i\varphi },\qquad \psi ^{\ast }=\sqrt{\rho }%
e^{-i\varphi }  \label{b1.14}
\end{equation}
to the action (\ref{b1.3}), we obtain the same result (\ref{b1.12}), (\ref
{b1.13}) with the constant $\hbar $ instead of the constant $b_{0}$.

Let us compare the dynamic disquantization (\ref{b1.8}) and the conventional
method of disquantization (when $\hbar \rightarrow 0$). The conventional
method is not formalized, and only given enough ingenuity, researchers can
apply it to new quantum systems, provided the result of disquantization is
known a priori. Besides, it refers to the quantum principles. On the
contrary, the dynamic disquantization is well defined and formalized. Any
literate student can apply this method for disquantization of any new
quantum system. He obtains an unique result without any a priori information
and without any reference to the quantum principles and quantum constant.

If we want to have a well defined disquantization procedure, we should
define it in the form of dynamic disquantization (\ref{b1.8}), but not by
means of the limit $\hbar \rightarrow 0$. The dynamic disquantization is not
a supposition, which should be founded, or tested. It is simply a method of
investigation of a continuous dynamic system $\mathcal{S}$, associating it
with a discrete dynamic system $\mathcal{S}_{\mathrm{cl}}$. Application of
this method to dynamic system $\mathcal{S}_{\mathrm{S}}$ (\ref{b1.3}) gives
the result $\mathcal{S}_{\mathrm{Scl}}$ (\ref{b1.6}), and it allows one to
interpret the procedure (\ref{b1.8}) as a disquantization. This
disquantization is dynamic, because it uses only properties of the dynamic
system $\mathcal{S}$ and nothing besides them.

We are forced to explain such evident things in details, because the first
version \cite{R2001} of this paper was rejected by several journals on
physics and mathematical physics and was not published. Referees of these
journals stated that the dynamic disquantization procedure is not
substantiated properly and results of this disquantization application to
the Dirac particle $\mathcal{S}_{\mathrm{D}}$ are not tested experimentally.
Opinion of the referees reflects the viewpoint of the scientific community,
and we are forced to explain situation despite absurdity of these objections.

Experimental test is necessary, if one makes some suppositions, and then the
experimental test shows, whether or not these suppositions are valid. We
make no suppositions. We obtain results by means of logical reasonings and
mathematical calculations. Experimental test of our results is the same as
an experimental test of the Newtonian binomial $\left( a+b\right)
^{2}=a^{2}+2ab+b^{2}$.

The dynamic disquantization (\ref{b1.8}) is not a supposition. It is a
definition of the procedure. One may test consistency of this procedure. One
may apply this procedure, or not apply it, but one may not demand
substantiation of the definition. One may consider motives of such a
definition, and a bit later we consider these motives, but these motives are
not a substantiation of the definition (\ref{b1.8}), and the definition does
not need any substantiation.

Before considerations of these motives, we try to answer the very important
question. Why the disquantization procedure has not been formalized? The
disquantization is a very important procedure. The dynamic disquantization (%
\ref{b1.8}) is very simple and evident procedure, but it has not been
discovered during eighty years of the quantum mechanics existence. Why? What
obstacles did prevent the disquantization from formalization?

The general answer is as follows. Researchers believed in quantum principles
and in quantum nature of the microcosm. They cannot imagine, that quantum
systems can be investigated without a use of quantum principles. Now
details. According to dominating Copenhagen interpretation the wave function 
$\psi $ of a quantum particle is a specific quantum object, which has not a
classical analog. The wave function is supposed to describe the state of
individual quantum particle. On the other hand, the wave function of an
individual quantum particle describes the state of a continuous dynamic
system, having infinite number of the freedom degrees. The classical
particle is described by a discrete dynamic system, having a finite number
of the freedom degrees. To formalize the disquantization procedure, one
needs to formalize the transition from the continuous dynamic system to the
discrete one. How can one formalize the jump from infinite number of the
freedom degrees to the finite one?

The problem is solved as follows. At first, one shows that the wave function
is not a specific quantum object. The wave function is a method of
description of any fluidlike continuous dynamic system \cite{R99}. Quantum
systems are dynamic systems of such a kind. But the pure statistical
ensembles $\mathcal{E}\left[ \mathcal{S}_{\mathrm{cl}}\right] $ of classical
systems $\mathcal{S}_{\mathrm{cl}}$ are also dynamic systems of such a kind.
The state of such an ensemble $\mathcal{E}\left[ \mathcal{S}_{\mathrm{cl}}%
\right] $ may be also described by the wave function. The classical system $%
\mathcal{S}_{\mathrm{cl}}$ and the statistical ensemble $\mathcal{E}\left[ 
\mathcal{S}_{\mathrm{cl}}\right] $ are coupled between themselves in the
sense, that the action $\mathcal{A}_{\mathcal{S}_{\mathrm{cl}}}$ for $%
\mathcal{S}_{\mathrm{cl}}$ determines the action $\mathcal{A}_{\mathcal{E}%
\left[ \mathcal{S}_{\mathrm{cl}}\right] }$ for $\mathcal{E}\left[ \mathcal{S}%
_{\mathrm{cl}}\right] $ and vice versa the action $\mathcal{A}_{\mathcal{E}%
\left[ \mathcal{S}_{\mathrm{cl}}\right] }$ determines the action $\mathcal{A}%
_{\mathcal{S}_{\mathrm{cl}}}$. Dynamic system $\mathcal{E}\left[ \mathcal{S}%
_{\mathrm{cl}}\right] $ is continuous, and it contains an infinite number of
the freedom degrees, whereas $\mathcal{S}_{\mathrm{cl}}$ is a discrete
dynamic system which contains a finite number of the freedom degrees.
Connection between $\mathcal{S}_{\mathrm{cl}}$ and $\mathcal{E}\left[ 
\mathcal{S}_{\mathrm{cl}}\right] $ allows one to overcome the jump between
the continuous dynamic system and the discrete one.

The difference between the quantum system $\mathcal{S}$ and the statistical
ensemble $\mathcal{E}\left[ \mathcal{S}_{\mathrm{cl}}\right] $ lies in the
form of dynamic equations. Dynamic equations for $\mathcal{S}$ are partial
differential equations, which cannot be transformed to the form of ordinary
differential equations, because they contain derivatives in different
directions, whereas dynamic equations for $\mathcal{E}\left[ \mathcal{S}_{%
\mathrm{cl}}\right] $ are ordinary differential equations, or partial
differential equations, which can be reduced to the ordinary differential
equations by means of a change of variables.

At the disquantization procedure all components of derivatives transversal
to the vector $j^{k}$ are suppressed, and the quantum system $\mathcal{S}$
turns into the statistical ensemble $\mathcal{E}\left[ \mathcal{S}_{\mathrm{%
cl}}\right] $. Having determined the statistical ensemble $\mathcal{E}\left[ 
\mathcal{S}_{\mathrm{cl}}\right] $, one can determine the classical dynamic
system $\mathcal{S}_{\mathrm{cl}}$. Of course, we must adopt that the
quantum dynamic system $\mathcal{S}$ is a statistical ensemble $\mathcal{E}%
\left[ \mathcal{S}_{\mathrm{st}}\right] $ of some individual stochastic
systems, but not an individual quantum system, because otherwise we cannot
explain, how individual quantum particle can turn into the statistical
ensemble $\mathcal{E}\left[ \mathcal{S}_{\mathrm{cl}}\right] $ of classical
particles. It means that the wave function describes the state of the
statistical ensemble of particles (classical, or quantum), but not an
individual particle. It means that the Copenhagen interpretation is false at
the point, when it states that the wave function describes the state of an
individual particle. This statement of the Copenhagen interpretation is
incompatible with the quantum mechanics formalism \cite{R2004}.

Note that our explanation of the situation with the formalization of the
disquantization procedure is purely dynamical. The difference between the
quantum system $\mathcal{S}$ and the statistical ensemble $\mathcal{E}\left[ 
\mathcal{S}_{\mathrm{cl}}\right] $ lies in the form of dynamic equations,
but not in enigmatic quantum principles. Hence, the problem of the
disquantization formalization is a dynamical problem, which should be solved
by dynamic methods.

Thus, there are three reasons, why the problem of the disquantization
formalization has not been solved: (1) belief that the wave function is a
specific quantum object, (2) belief that the wave function describes the
state of individual quantum particle, (3) belief in principles of quantum
mechanics, and attempts to solve the problem on their basis.

Now about physical reasons of the dynamic disquantization (\ref{b1.8}).
Conventional approach to disquantization is an attempt of cutting off the
quantum stochasticity, setting $\hbar =0$ in the proper representation (\ref
{b1.5}) of the action for $\mathcal{S}_{\mathrm{S}}$. Dynamic
disquantization does not try to cut off the quantum stochasticity. It uses
existence of such states of the statistical ensemble of stochastic systems $%
\mathcal{S}_{\mathrm{st}}$, where the stochastic component of the particle
motion does not influence upon the regular component. Stochasticity
influences upon the regular component only in nonuniform states. This
statement is valid for all stochastic systems (but not only for quantum
ones). For instance, the mean velocity of Brownian particles is determined
by the relation 
\[
\mathbf{v}_{\mathrm{B}}=-D\mathbf{\nabla }\text{ln}\rho 
\]
where $\rho $ is the Brownian particle density, and $D$ is the diffusion
coefficient. If the ensemble of Brownian particles is uniform, $\rho =$%
const, the mean velocity vanishes, although the random motion of Brownian
particles remains. Analogously, if in the statistical ensemble $\mathcal{E}%
\left[ \mathcal{S}_{\mathrm{st}}\right] $ $\;\;\mathbf{\nabla }u=0$ for all
physical quantities $u$, the influence of the stochastic component on the
regular component vanishes. The space gradient $\mathbf{\nabla }u$ is
considered in the coordinate system, where the medium is at rest, and the
current vector $j^{k}$ has the form $j^{k}=\left\{ j^{0},0,0,0\right\} $. In
the arbitrary coordinate system the condition $\mathbf{\nabla }u=0$ turns
into 
\begin{equation}
\partial _{\bot }^{k}u=\partial ^{k}u-\frac{j^{k}j^{l}}{j^{s}j_{s}}\partial
_{l}=0  \label{b1.15}
\end{equation}
which can be realized by means of the change (\ref{b1.8}). This
consideration is only an \textit{explanation of the dynamic disquantization,
but not its substantiation}.

Further we transform the action (\ref{b1.1}) for the Dirac particle $%
\mathcal{S}_{\mathrm{D}}$ to hydrodynamical variables, where the current
components $j^{k}=\bar{\psi}\gamma ^{k}\psi $,\ $k=0,1,2,3$ are dependent
variables. We produce dynamic disquantization (\ref{b1.8}) and obtain
classical dynamic system $\mathcal{S}_{\mathrm{Dcl}}$, having ten degrees of
freedom. We solve the dynamic equations for $\mathcal{S}_{\mathrm{Dcl}}$ and
find that $\mathcal{S}_{\mathrm{Dcl}}$ can be identified with a rotator
which also has ten degrees of freedom.

The goal of investigation is a construction of dynamic system $\mathcal{S}_{%
\mathrm{Dcl}}$, associated with $\mathcal{S}_{\mathrm{D}}$. Dynamic
equations for $\mathcal{S}_{\mathrm{Dcl}}$ form a system of ordinary
differential equations. Further the dynamic system $\mathcal{S}_{\mathrm{Dcl}%
}$, will be referred to as the classical Dirac particle. It has finite
number of the freedom degrees, and it is simpler for investigation, than $%
\mathcal{S}_{\mathrm{D}}$.

\textit{If our statement on the composite structure of the Dirac particle
appears to be incompatible with experimental data, this result is an
argument against application of the Dirac equation, but not against our
investigation of the Dirac equation.}

Investigating the Dirac particle $\mathcal{S}_{\mathrm{D}}$, we transform
the action (\ref{b1.1}) to hydrodynamic variables by means of a change of
variables. In terms of the new variables the variables $j^{k}$, defined by (%
\ref{f1.3}) are four dependent variables. Form of other four dependent
variables is chosen in such a way, to eliminate $\gamma $-matrices from the
action. After such a change of variables we can produce the dynamic
disquantization in the action, making the change (\ref{b1.8}). Thereafter
the action for $\mathcal{S}_{\mathrm{D}}$ turns into the action for the
statistical ensemble of classical particles $\mathcal{S}_{\mathrm{Dcl}}$,
having ten degrees of freedom. Investigating properties of classical dynamic
system $\mathcal{S}_{\mathrm{Dcl}}$, we investigate properties of the Dirac
particle $\mathcal{S}_{\mathrm{D}}$. This investigation allows one to
interpret properties of the Dirac particle $\mathcal{S}_{\mathrm{D}}$ in
terms of the classical dynamic system $\mathcal{S}_{\mathrm{Dcl}}$.

\section{Transformation of variables}

The state of dynamic system $\mathcal{S}_{\mathrm{D}}$ is described by eight
real dependent variables (eight real components of four-component complex
wave function $\psi $). Transforming the action (\ref{b1.1}), we use the
mathematical technique \cite{S30,S51}, where the wave function $\psi $ is
considered to be a function of hypercomplex numbers $\gamma $ and
coordinates $x$. In this case the dynamical quantities are obtained by means
of a convolution of expressions $\psi ^{\ast }O\psi $ with zero divisors.
This technique allows one to work without fixing the $\gamma $-matrices
representation.

Using designations 
\begin{equation}
\gamma _{5}=\gamma ^{0123}\equiv \gamma ^{0}\gamma ^{1}\gamma ^{2}\gamma
^{3},  \label{f1.9}
\end{equation}
\begin{equation}
\mathbf{\sigma }=\{\sigma _{1},\sigma _{2},\sigma _{3},\}=\{-i\gamma
^{2}\gamma ^{3},-i\gamma ^{3}\gamma ^{1},-i\gamma ^{1}\gamma ^{2}\}
\label{f1.10}
\end{equation}
we make the change of variables 
\begin{equation}
\psi =Ae^{i\varphi +{\frac{1}{2}}\gamma _{5}\kappa }\exp \left( -\frac{i}{2}%
\gamma _{5}\mathbf{\sigma \eta }\right) \exp \left( {\frac{i\pi }{2}}\mathbf{%
\sigma n}\right) \Pi  \label{f1.11}
\end{equation}
\begin{equation}
\psi ^{\ast }=A\Pi \exp \left( -{\frac{i\pi }{2}}\mathbf{\sigma n}\right)
\exp \left( -\frac{i}{2}\gamma _{5}\mathbf{\sigma \eta }\right) e^{-i\varphi
-{\frac{1}{2}}\gamma _{5}\kappa }  \label{f1.12}
\end{equation}
where (*) means the Hermitian conjugation, and 
\begin{equation}
\Pi ={\frac{1}{4}}(1+\gamma ^{0})(1+\mathbf{z\sigma }),\qquad \mathbf{z}%
=\{z^{\alpha }\}=\text{const},\qquad \alpha =1,2,3;\qquad \mathbf{z}^{2}=1
\label{f1.13}
\end{equation}
is a zero divisor. The quantities $A$, $\kappa $, $\varphi $, $\mathbf{\eta }%
=\{\eta ^{\alpha }\}$, $\mathbf{n}=\{n^{\alpha }\}$, $\alpha =1,2,3,\;$ $%
\mathbf{n}^{2}=1$ are eight real parameters, determining the wave function $%
\psi .$ These parameters may be considered as new dependent variables,
describing the state of dynamic system $\mathcal{S}_{\mathrm{D}}$. The
quantity $\varphi $ is a scalar, and $\kappa $ is a pseudoscalar. Six
remaining variables $A,$ $\mathbf{\eta }=\{\eta ^{\alpha }\}$, $\mathbf{n}%
=\{n^{\alpha }\}$, $\alpha =1,2,3,\;$ $\mathbf{n}^{2}=1$ can be expressed
through the flux 4-vector $j^{l}=\bar{\psi}\gamma ^{l}\psi $ and spin
4-pseudovector 
\begin{equation}
S^{l}=i\bar{\psi}\gamma _{5}\gamma ^{l}\psi ,\qquad l=0,1,2,3  \label{f1.13a}
\end{equation}
Because of two identities 
\begin{equation}
S^{l}S_{l}\equiv -j^{l}j_{l},\qquad j^{l}S_{l}\equiv 0.  \label{f1.14}
\end{equation}
there are only six independent components among eight components of
quantities $j^{l}$, and $S^{l}$. .

Matrices $\gamma _{5}$, $\mathbf{\sigma }=\{\sigma _{\alpha }\},$ $\alpha
=1,2,3$ are determined by relations (\ref{f1.9}), (\ref{f1.10}) have the
following properties 
\begin{equation}
\gamma _{5}\gamma _{5}=-1,\qquad \gamma _{5}\sigma _{\alpha }=\sigma
_{\alpha }\gamma _{5},\qquad \gamma ^{0\alpha }\equiv \gamma ^{0}\gamma
^{\alpha }=-i\gamma _{5}\sigma _{\alpha },\qquad \alpha =1,2,3;  \label{b4.4}
\end{equation}
\begin{equation}
\left( \gamma ^{0}\right) ^{\ast }=\gamma ^{0},\qquad \left( \gamma ^{\alpha
}\right) ^{\ast }=-\gamma ^{\alpha },\qquad \gamma ^{0}\mathbf{\sigma }=%
\mathbf{\sigma }\gamma ^{0},\qquad \gamma ^{0}\gamma _{5}=-\gamma _{5}\gamma
^{0}  \label{b4.5}
\end{equation}
According to relations (\ref{b1.2}), (\ref{f1.9}), (\ref{f1.10}) the
matrices $\mathbf{\sigma }=\{\sigma _{\alpha }\}$, $\alpha =1,2,3$ satisfy
the relation 
\begin{equation}
\sigma _{\alpha }\sigma _{\beta }=\delta _{\alpha \beta }+i\varepsilon
_{\alpha \beta \gamma }\sigma _{\gamma },\qquad \alpha ,\beta =1,2,3
\label{a3.3}
\end{equation}
where $\varepsilon _{\alpha \beta \gamma }$ is the antisymmetric
pseudo-tensor of Levi-Chivita $(\varepsilon _{123}=1)$.

Using relations (\ref{b4.4}),(\ref{b4.5}), (\ref{a3.3}) and (\ref{f1.13}),
it is easy to verify that 
\begin{eqnarray}
\Pi ^{2} &=&\Pi ,\qquad \gamma _{0}\Pi =\Pi ,\qquad \mathbf{z\sigma }\Pi
=\Pi ,  \label{b4.5a} \\
\Pi \gamma _{5}\Pi &=&0,\qquad \Pi \sigma _{\alpha }\Pi =z^{\alpha }\Pi
,\qquad \alpha =1,2,3.  \label{a3.6}
\end{eqnarray}
Generally, the wave functions $\psi ,\psi ^{\ast }$ defined by (\ref{f1.12})
are $4\times 4$ complex matrices. In the proper representation, where $\Pi $
has the form 
\begin{equation}
\Pi =\left( 
\begin{array}{cccc}
1 & 0 & 0 & 0 \\ 
0 & 0 & 0 & 0 \\ 
0 & 0 & 0 & 0 \\ 
0 & 0 & 0 & 0
\end{array}
\right)  \label{a3.7}
\end{equation}
the $\psi ,\psi ^{\ast }$ have the form

\begin{equation}
\psi =\left( 
\begin{array}{cccc}
\psi _{1} & 0 & 0 & 0 \\ 
\psi _{2} & 0 & 0 & 0 \\ 
\psi _{3} & 0 & 0 & 0 \\ 
\psi _{4} & 0 & 0 & 0
\end{array}
\right) ,\qquad \psi ^{\ast }=\left( 
\begin{array}{cccc}
\psi _{1}^{\ast } & \psi _{2}^{\ast } & \psi _{3}^{\ast } & \psi _{4}^{\ast }
\\ 
0 & 0 & 0 & 0 \\ 
0 & 0 & 0 & 0 \\ 
0 & 0 & 0 & 0
\end{array}
\right)  \label{a3.8}
\end{equation}
Let $O$ be an arbitrary $4\times 4$ matrix. The product $\psi ^{\ast }O\psi $
has the form 
\begin{equation}
\psi ^{\ast }O\psi =\left( 
\begin{array}{cccc}
a & 0 & 0 & 0 \\ 
0 & 0 & 0 & 0 \\ 
0 & 0 & 0 & 0 \\ 
0 & 0 & 0 & 0
\end{array}
\right) =a\Pi =\Pi a  \label{a3.9}
\end{equation}
where $a$ is a complex quantity. If $f$ is an analytical function having the
property $f(0)=0,$ then the function $f(\psi ^{\ast }O\psi )=f(a\Pi )$ of a $%
4\times 4$ matrix of the type (\ref{a3.9}) is a matrix $f(a)\Pi $ of the
same type. For this reason we shall not distinguish between the complex
quantity $a$ and the complex $4\times 4$ matrix $a\Pi $. In the final
expressions of the type $a\Pi $ ($a$ is a complex quantity) the multiplier $%
\Pi $ will be omitted.

By means of relations (\ref{b4.4}) -- (\ref{a3.6}), one can reduce any
Clifford number $\Pi O\Pi $ to the form (\ref{a3.9}), without using any
concrete form of the $\gamma $-matrix representation. This property will be
used in our calculations. Calculating exponents of the type (\ref{f1.11}), (%
\ref{f1.12}), we shall use the following relations 
\[
\exp \left( -{\frac{i\pi }{2}}\mathbf{\sigma n}\right) F\left( \mathbf{%
\sigma }\right) \exp \left( {\frac{i\pi }{2}}\mathbf{\sigma n}\right)
=F\left( \mathbf{\Sigma }\right) 
\]
where $F$ is arbitrary function and the quantity 
\begin{equation}
\mathbf{\Sigma }=\{\Sigma _{1},\Sigma _{2},\Sigma _{3}\},\qquad \Sigma
_{\alpha }=\exp \left( -{\frac{i\pi }{2}}\mathbf{\sigma n}\right) \sigma
_{\alpha }\exp \left( {\frac{i\pi }{2}}\mathbf{\sigma n}\right) \qquad
\alpha =1,2,3;  \label{a3.12}
\end{equation}
satisfies the same commutation relations (\ref{a3.3}) as the Pauli matrices $%
\mathbf{\sigma }$.

For variables $\bar{\psi}\psi $, $j^{l}$, $S^{l}$, $l=0,1,2,3$ we have the
following expressions 
\[
\bar{\psi}\psi =\psi ^{\ast }\gamma ^{0}\psi =A^{2}\Pi e^{\gamma _{5}\kappa
}\Pi =A^{2}\Pi \left( \cos \kappa +\gamma _{5}\sin \kappa \right) \Pi
=A^{2}\cos \kappa \Pi 
\]
Taking into account the first relation (\ref{a3.6}), the term linear with
respect to $\gamma _{5}$ vanishes, and we obtain 
\[
\bar{\psi}\psi =A^{2}\cos \kappa \Pi 
\]
\begin{eqnarray}
j^{0}\Pi &=&\bar{\psi}\gamma ^{0}\psi =A^{2}\Pi \exp \left( -{\frac{i\pi }{2}%
}\mathbf{\sigma n}\right) \exp \left( -i\gamma _{5}\mathbf{\sigma \eta }%
\right) \exp \left( {\frac{i\pi }{2}}\mathbf{\sigma n}\right) \Pi  \nonumber
\\
&=&A^{2}\Pi \exp \left( -i\gamma _{5}\mathbf{\Sigma \eta }\right) \Pi
=A^{2}\Pi \left( \cosh \eta -\frac{i\gamma _{5}}{\eta }\mathbf{\Sigma \eta }%
\sinh \eta \right) \Pi  \nonumber \\
&=&A^{2}\cosh (\eta )\Pi  \label{f1.45}
\end{eqnarray}
where 
\[
\eta =\sqrt{\mathbf{\eta }^{2}}=\sqrt{\eta ^{\alpha }\eta ^{\alpha }} 
\]
Again in force of the first relation (\ref{a3.6}) we omit terms linear with
respect to $\gamma _{5}$.

In the same way we obtain 
\begin{eqnarray*}
j^{\alpha }\Pi &=&\psi ^{\ast }\gamma ^{0\alpha }\psi \Pi =A^{2}\Pi \exp
\left( -\frac{i}{2}\gamma _{5}\mathbf{\Sigma \eta }\right) (-i\gamma
_{5}\Sigma _{\alpha })\exp \left( -\frac{i}{2}\gamma _{5}\mathbf{\Sigma \eta 
}\right) \Pi = \\
&=&A^{2}\Pi (\cosh \frac{\eta }{2}-i\gamma _{5}\mathbf{\Sigma v}\sinh \frac{%
\eta }{2})(-i\gamma _{5}\Sigma _{\alpha })(\cosh \frac{\eta }{2}-i\gamma _{5}%
\mathbf{\Sigma v}\sinh \frac{\eta }{2})\Pi = \\
&=&A^{2}\Pi (\cosh \frac{\eta }{2}\sinh \frac{\eta }{2}\left( \Sigma _{\beta
}\Sigma _{\alpha }+\Sigma _{\beta }\Sigma _{\alpha }\right) v^{\beta }\Pi
\end{eqnarray*}
\begin{equation}
j^{\alpha }\Pi =A^{2}\sinh (\eta )v^{\alpha }\Pi ,\qquad \alpha =1,2,3
\label{a3.13}
\end{equation}
where 
\begin{equation}
\mathbf{v}=\{v^{\alpha }\},\qquad v^{\alpha }=\eta ^{\alpha }/\eta ,\qquad
\alpha =1,2,3;\qquad \mathbf{v}^{2}=1.  \label{a3.14}
\end{equation}

Let us introduce designation $\mathbf{\xi }=\{\xi ^{\alpha }\}$, $\alpha
=1,2,3$ for the expression 
\begin{equation}
\xi ^{\alpha }\Pi =\Pi \Sigma _{\alpha }\Pi ,\qquad \alpha =1,2,3,\qquad 
\mathbf{\xi }^{2}=\xi ^{\alpha }\xi ^{\alpha }=1  \label{a3.16}
\end{equation}
Then for the spin pseudovector $S^{l}$, defined by the relation (\ref{f1.13a}%
), we obtain

\begin{equation}
\begin{array}{lll}
S^{0}\Pi & = & \psi ^{\ast }(-i\gamma _{5})\psi =A^{2}\Pi (-i\gamma
_{5})\exp \left( -i\gamma _{5}\mathbf{\Sigma \eta }\right) \Pi = \\ 
& = & A^{2}\Pi \sinh (\eta )\mathbf{\Sigma }\mathbf{v}\Pi =A^{2}\sinh (\eta )%
\mathbf{\xi v}\Pi ,
\end{array}
\label{a3.14a}
\end{equation}
\begin{eqnarray*}
S^{\alpha }\Pi &=&\psi ^{\ast }\gamma ^{0}i\gamma _{5}\gamma ^{\alpha }\psi
=\Pi \psi ^{\ast }\sigma _{\alpha }\psi \Pi =A^{2}\Pi \exp \left( -\frac{i}{2%
}\gamma _{5}\mathbf{\Sigma \eta }\right) \Sigma _{\alpha }\exp \left( -\frac{%
i}{2}\gamma _{5}\mathbf{\Sigma \eta }\right) \Pi = \\
&=&A^{2}\Pi (\cosh {\frac{\eta }{2}}-i\gamma _{5}\mathbf{\Sigma v}\sinh {%
\frac{\eta }{2}})\Sigma _{\alpha }(\cosh {\frac{\eta }{2}}-i\gamma _{5}%
\mathbf{\Sigma v}\sinh {\frac{\eta }{2}})\Pi = \\
&=&A^{2}\Pi \left( \cosh ^{2}\frac{\eta }{2}\Sigma _{\alpha }+\sinh ^{2}{%
\frac{\eta }{2}}\left( \Sigma _{\beta }v^{\beta }\right) \Sigma _{\alpha
}\left( \Sigma _{\gamma }v^{\gamma }\right) \right) \Pi
\end{eqnarray*}
Now twice using relations (\ref{a3.3}) for Pauli matrices $\Sigma _{\alpha }$%
, we derive 
\begin{eqnarray*}
S^{\alpha }\Pi &=&A^{2}\Pi \left( \cosh ^{2}\frac{\eta }{2}\Sigma _{\alpha
}+\sinh ^{2}{\frac{\eta }{2}}v^{\beta }v^{\gamma }\left( \delta _{\alpha
\beta }+i\varepsilon _{\beta \alpha \mu }\Sigma _{\mu }\right) \Sigma
_{\gamma }\right) \Pi = \\
&=&A^{2}\Pi \left( \cosh ^{2}\frac{\eta }{2}\Sigma _{\alpha }+\sinh ^{2}{%
\frac{\eta }{2}}\left( v^{\alpha }v^{\gamma }\Sigma _{\gamma }+i\varepsilon
_{\beta \alpha \mu }v^{\beta }v^{\gamma }\left( \delta _{\mu \gamma
}+i\varepsilon _{\mu \gamma \nu }\Sigma _{\nu }\right) \right) \right) \Pi \\
&=&A^{2}\Pi \left( \cosh ^{2}\frac{\eta }{2}\Sigma _{\alpha }+\sinh ^{2}{%
\frac{\eta }{2}}\left( v^{\alpha }v^{\gamma }\Sigma _{\gamma }-v^{\beta
}v^{\beta }\Sigma _{\alpha }+v^{\beta }v^{\alpha }\Sigma _{\beta }\right)
\right) \Pi
\end{eqnarray*}
\begin{equation}
S^{\alpha }\Pi =A^{2}[\xi ^{\alpha }+(\cosh \eta -1)v^{\alpha }(\mathbf{v\xi 
})]\Pi ,\qquad \alpha =1,2,3.  \label{a3.15}
\end{equation}

It follows from relations (\ref{f1.45}), (\ref{a3.13}), (\ref{a3.14}) 
\begin{equation}
j^{i}j_{i}\Pi =A^{4}\Pi ,\qquad A=(j^{l}j_{l})^{1/4}\equiv \rho ^{1/2}
\label{a3.17}
\end{equation}
According to the third equation (\ref{b4.5a}), (\ref{a3.12}) and (\ref{a3.16}%
) one obtains 
\begin{eqnarray*}
\xi ^{\alpha }\Pi &=&\Pi \Sigma _{\alpha }\Pi =\Pi \exp \left( -\frac{i\pi }{%
2}\mathbf{\sigma n}\right) \sigma _{\alpha }\exp \left( \frac{i\pi }{2}%
\mathbf{\sigma n}\right) \Pi = \\
&=&\Pi \left( \cos \frac{\pi }{2}-i\mathbf{\sigma n}\sin \frac{\pi }{2}%
\right) \sigma _{\alpha }\left( \cos \frac{\pi }{2}+i\mathbf{\sigma n}\sin 
\frac{\pi }{2}\right) \Pi = \\
&=&\Pi \left( \mathbf{\sigma n}\right) \sigma _{\alpha }\left( \mathbf{%
\sigma n}\right) \Pi =\Pi n^{\mu }n^{\nu }\sigma _{\mu }\sigma _{\alpha
}\sigma _{\nu }\Pi = \\
&=&\Pi \left( n^{\alpha }n^{\nu }\sigma _{\nu }+i\varepsilon _{\mu \alpha
\gamma }\sigma _{\gamma }\sigma _{\nu }n^{\mu }n^{\nu }\right) \Pi =\Pi
\left( n^{\alpha }n^{\nu }\sigma _{\nu }-\varepsilon _{\mu \alpha \gamma
}\varepsilon _{\gamma \nu \beta }\sigma _{\beta }n^{\mu }n^{\nu }\right) \Pi
= \\
&=&\Pi \left( n^{\alpha }n^{\nu }z_{\nu }-\varepsilon _{\mu \alpha \gamma
}\varepsilon _{\gamma \nu \beta }z^{\beta }n^{\mu }n^{\nu }\right) \Pi \\
&=&\left( n^{\alpha }\left( \mathbf{nz}\right) +\left( \mathbf{n}\times
\left( \mathbf{n}\times \mathbf{z}\right) \right) ^{\alpha }\right) \Pi
,\qquad \alpha =1,2,3;
\end{eqnarray*}
Or

\begin{equation}
\mathbf{\xi }=2\mathbf{n}(\mathbf{nz})-\mathbf{z}  \label{a3.21}
\end{equation}
where $\mathbf{z}$ is defined by (\ref{f1.13}).

\section{Transformation of the action}

Let us make a change of variables in the action (\ref{b1.1}), using
substitution (\ref{f1.11}) -- (\ref{f1.13}). The last two terms of the
action (\ref{b1.1}) may be written in the form 
\begin{eqnarray*}
{\frac{i}{2}}\hbar \bar{\psi}\gamma ^{l}\partial _{l}\psi +\text{h.c} &=&{%
\frac{i}{2}}\hbar \psi ^{\ast }\left( \partial _{0}-i\gamma _{5}\mathbf{%
\sigma \nabla }\right) \psi +\text{h.c} \\
&=&{\frac{i}{2}}\hbar \psi ^{\ast }\left( \left( \partial _{0}-i\gamma _{5}%
\mathbf{\sigma \nabla }\right) \left( i\varphi +\frac{1}{2}\gamma _{5}\kappa
\right) \right) \psi +\text{h.c.} \\
&&+\frac{i}{2}\hbar A^{2}\Pi \exp \left( -\frac{i\pi }{2}\mathbf{\sigma n}%
\right) \exp \left( -\frac{i}{2}\gamma _{5}\mathbf{\Sigma \eta }\right)
(\partial _{0}-i\gamma _{5}\mathbf{\sigma \nabla }) \\
&&\times (\exp \left( -\frac{i}{2}\gamma _{5}\mathbf{\Sigma \eta }\right)
\exp \left( \frac{i\pi }{2}\mathbf{\sigma n}\right) )\Pi +\text{h.c}
\end{eqnarray*}
where ''h.c.'' means the term obtained from the previous one by the
Hermitian conjugation. Calculation of this expression gives the following
result (see details of calculation in Appendix A). 
\begin{equation}
{\frac{i}{2}}\hbar \bar{\psi}\gamma ^{l}\partial _{l}\psi +\text{h.c}%
=F_{1}+F_{2}+F_{3}+F_{4}  \label{a4.1}
\end{equation}
where 
\begin{equation}
F_{1}+F_{2}=-j^{l}\partial _{l}\varphi \Pi -{\frac{1}{2}}\hbar S^{l}\partial
_{l}\kappa \Pi  \label{b5.5a}
\end{equation}
\begin{equation}
F_{3}=-\frac{\hbar j^{l}}{2\left( 1+\mathbf{\xi z}\right) }\varepsilon
_{\alpha \beta \gamma }\xi ^{\alpha }\partial _{l}\xi ^{\beta }z^{\gamma }\Pi
\label{b5.10}
\end{equation}
\begin{equation}
F_{4}=\left( \frac{\hbar (\rho +j_{0})}{2}\varepsilon _{\alpha \beta \gamma
}\partial ^{\alpha }\frac{j^{\beta }}{(j^{0}+\rho )}\xi ^{\gamma }-\frac{%
\hbar }{2(\rho +j_{0})}\varepsilon _{\alpha \beta \gamma }\left( \partial
^{0}j^{\beta }\right) j^{\alpha }\xi ^{\gamma }\right) \Pi  \label{a4.6}
\end{equation}
Here $\varepsilon _{\alpha \beta \gamma }$ is 3-dimensional Levi-Chivita
pseudotensor.

We see that the expressions (\ref{b5.5}) for $F_{1}$ and $F_{2}$ as well as
the first term of the action (\ref{b1.1}) 
\begin{equation}
-m\bar{\psi}\psi =-m\Pi e^{\gamma _{5}\kappa }\Pi =-mA^{2}\cos \kappa \Pi =-m%
\sqrt{j^{l}j_{l}}\cos \kappa \Pi \equiv -m\rho \cos \kappa \Pi  \label{b5.11}
\end{equation}
have relativistically covariant form. The terms $F_{3}$ and $F_{4}$ have
non-covariant form. Introducing the constant unit 4-vector $f^{k}=\left\{
1,0,0,0\right\} $, they can be written in the relativistically covariant
form (see \cite{R2001}). The constant 4-vector $f^{k}$ appears from the
matrix 4-vector $\gamma ^{k}$, $k=0,1,2,3$, which figures in the original
action (\ref{b1.1})

Now we can write the action (\ref{b1.1}) in the hydrodynamical form 
\begin{equation}
\mathcal{S}_{\mathrm{D}}:\qquad \mathcal{A}_{D}[j,\varphi ,\kappa ,\mathbf{%
\xi }]=\int \mathcal{L}d^{4}x,\qquad \mathcal{L}=\mathcal{L}_{\mathrm{cl}}+%
\mathcal{L}_{\mathrm{q1}}+\mathcal{L}_{\mathrm{q2}}  \label{c4.15}
\end{equation}
\begin{equation}
\mathcal{L}_{\mathrm{cl}}=-m\rho -\hbar j^{i}\partial _{i}\varphi -\frac{%
\hbar j^{l}}{2\left( 1+\mathbf{\xi z}\right) }\varepsilon _{\alpha \beta
\gamma }\xi ^{\alpha }\partial _{l}\xi ^{\beta }z^{\gamma },\qquad \rho
\equiv \sqrt{j^{l}j_{l}}  \label{c4.16}
\end{equation}
\begin{equation}
\mathcal{L}_{\mathrm{q1}}=2m\rho \sin ^{2}(\frac{\kappa }{2})-{\frac{\hbar }{%
2}}S^{l}\partial _{l}\kappa ,  \label{c4.17}
\end{equation}
\begin{equation}
\mathcal{L}_{\mathrm{q2}}=\frac{\hbar (\rho +j_{0})}{2}\varepsilon _{\alpha
\beta \gamma }\partial ^{\alpha }\frac{j^{\beta }}{(j^{0}+\rho )}\xi
^{\gamma }-\frac{\hbar }{2(\rho +j_{0})}\varepsilon _{\alpha \beta \gamma
}\left( \partial ^{0}j^{\beta }\right) j^{\alpha }\xi ^{\gamma }
\label{c4.18}
\end{equation}
Lagrangian is a function of 4-vector $j^{l}$, scalar $\varphi $,
pseudoscalar $\kappa $, and unit 3-pseudovector $\mathbf{\xi }$, which is
connected with the spin 4-pseudovector $S^{l}$ by means of the relations 
\begin{equation}
\xi ^{\alpha }=\rho ^{-1}\left[ S^{\alpha }-\frac{j^{\alpha }S^{0}}{%
(j^{0}+\rho )}\right] ,\qquad \alpha =1,2,3;\qquad \rho \equiv \sqrt{%
j^{l}j_{l}}  \label{f1.15}
\end{equation}
\begin{equation}
S^{0}=\mathbf{j\xi },\qquad S^{\alpha }=\rho \xi ^{\alpha }+\frac{(\mathbf{%
j\xi })j^{\alpha }}{\rho +j^{0}},\qquad \alpha =1,2,3  \label{f1.16}
\end{equation}

\section{Dynamic disquantization}

Let us produce dynamical disquantization of the action (\ref{c4.15})--(\ref
{c4.18}), making the change (\ref{b1.8}). The action (\ref{c4.15})--(\ref
{c4.18}) takes the form 
\begin{eqnarray}
\mathcal{A}_{\mathrm{Dqu}}[j,\varphi ,\kappa ,\mathbf{\xi }] &=&\int \left\{
-m\rho \cos \kappa -\hbar j^{i}\left( \partial _{i}\varphi +\frac{%
\varepsilon _{\alpha \beta \gamma }\xi ^{\alpha }\partial _{i}\xi ^{\beta
}z^{\gamma }}{2\left( 1+\mathbf{\xi z}\right) }\right) \right.  \nonumber \\
&&+\left. \frac{\hbar j^{k}}{2(\rho +j_{0})\rho }\varepsilon _{\alpha \beta
\gamma }\left( \partial _{k}j^{\beta }\right) j^{\alpha }\xi ^{\gamma
}\right\} d^{4}x  \label{a5.9}
\end{eqnarray}
Note that the second term $-\frac{\hbar }{2}S^{l}\partial _{l}\kappa $ in
the relation (\ref{c4.17}) is neglected, because 4-pseudovector $S^{k}$ is
orthogonal to 4-vector $j^{k}$, and the derivative $S^{l}\partial
_{||l}\kappa =S^{l}\rho ^{-2}j_{l}j^{k}\partial _{k}\kappa $ vanishes.

Although the action (\ref{a5.9}) contains a non-classical variable $\kappa $%
, but in fact $\kappa $ is a constant quantity.. Indeed, a variation with
respect to $\kappa $ leads to the dynamic equation 
\begin{equation}
\frac{\delta \mathcal{A}_{\mathrm{Dqu}}}{\delta \kappa }=m\rho \sin \kappa =0
\label{a5.10}
\end{equation}
which has solutions 
\begin{equation}
\kappa =n\pi  \label{a5.11}
\end{equation}
where $n$ is integer. Thus, the effective mass $m_{\mathrm{eff}}=m\cos
\kappa $ has two values 
\begin{equation}
m_{\mathrm{eff}}=m\cos \kappa =\kappa _{0}m  \label{a5.12}
\end{equation}
where $\kappa _{0}$ is a dichotomic quantity $\kappa _{0}=\pm 1$ introduced
instead of cos $\kappa $. The quantity $\kappa _{0}$ is a parameter of the
dynamic system $\mathcal{S}_{\mathrm{Dqu}}$. It is not to be varying. The
action (\ref{a5.9}), turns into the action 
\begin{eqnarray}
\mathcal{A}_{\mathrm{Dqu}}[j,\varphi ,\mathbf{\xi }] &=&\int \left\{ -\kappa
_{0}m\rho -\hbar j^{i}\left( \partial _{i}\varphi +\frac{\varepsilon
_{\alpha \beta \gamma }\xi ^{\alpha }\partial _{i}\xi ^{\beta }z^{\gamma }}{%
2\left( 1+\mathbf{\xi z}\right) }\right) \right.  \nonumber \\
&&+\left. \frac{\hbar j^{k}}{2(\rho +j_{0})\rho }\varepsilon _{\alpha \beta
\gamma }\left( \partial _{k}j^{\beta }\right) j^{\alpha }\xi ^{\gamma
}\right\} d^{4}x  \label{a5.13a}
\end{eqnarray}

Let us introduce Lagrangian coordinates $\tau =\{\tau _{0},\mathbf{\tau }%
\}=\{\tau _{i}\left( x\right) \}$, $i=0,1,2,3$ as functions of coordinates $%
x $ in such a way that only coordinate $\tau _{0}$ changes along the
direction $j^{l},$ i.e. 
\begin{equation}
j^{k}\partial _{k}\tau _{\mu }=0,\qquad \mu =1,2,3  \label{b3.1}
\end{equation}
Considering coordinates $x$ to be a functions of $\tau =\left\{ \tau _{0},%
\mathbf{\tau }\right\} $, one has the following identities 
\begin{equation}
\frac{\partial D}{\partial \tau _{0,k}}\tau _{i,k}\equiv \delta
_{i}^{0}D,\qquad i=0,1,2,3\qquad \tau _{i,k}\equiv \partial _{k}\tau
_{i},\qquad i,k=0,1,2,3  \label{b3.2}
\end{equation}
where 
\begin{equation}
D\equiv \frac{\partial (\tau _{0},\tau _{1},\tau _{2},\tau _{3})}{\partial
(x^{0},x^{1},x^{2},x^{3})},\qquad \frac{\partial D}{\partial \tau _{0,i}}%
\equiv \frac{\partial (x^{i},\tau _{1},\tau _{2},\tau _{3})}{\partial
(x^{0},x^{1},x^{2},x^{3})}.  \label{a5.14}
\end{equation}

Comparing (\ref{b3.1}) with (\ref{b3.2}), one concludes that it is possible
to set 
\begin{equation}
j^{i}=\frac{\partial D}{\partial \tau _{0,i}}\equiv \frac{\partial
(x^{i},\tau _{1},\tau _{2},\tau _{3})}{\partial (x^{0},x^{1},x^{2},x^{3})}%
,\qquad i=0,1,2,3  \label{b3.3}
\end{equation}
because the dynamic equation 
\begin{equation}
\frac{\delta \mathcal{A}_{\mathrm{Dqu}}}{\delta \varphi }=\hbar \partial
_{l}j^{l}=0  \label{b3.4}
\end{equation}
is satisfied by the relation (\ref{b3.3}) identically in force of identity 
\[
\partial _{i}\frac{\partial D}{\partial \tau _{k,i}}\equiv 0,\qquad
k=0,1,2,3. 
\]

Let us take into account that for any variable $u$%
\begin{equation}
D^{-1}j^{i}\partial _{i}u=D^{-1}\frac{\partial D}{\partial \tau _{0,i}}%
\partial _{i}u=\frac{\partial (u,\tau _{1},\tau _{2},\tau _{3})}{\partial
(\tau _{0},\tau _{1},\tau _{2},\tau _{3})}=\frac{du}{d\tau _{0}}
\label{b3.6}
\end{equation}
and in particular, 
\begin{equation}
D^{-1}j^{i}=D^{-1}\frac{\partial D}{\partial \tau _{0,i}}\equiv \frac{%
\partial (x^{i},\tau _{1},\tau _{2},\tau _{3})}{\partial (\tau _{0},\tau
_{1},\tau _{2},\tau _{3})}=\frac{dx^{i}}{d\tau _{0}}\equiv \dot{x}%
^{i},\qquad i=0,1,2,3  \label{b3.7}
\end{equation}
Besides 
\begin{equation}
d^{4}x=D^{-1}d^{4}\tau =D^{-1}d\tau _{0}d\mathbf{\tau }  \label{a5.16}
\end{equation}
\begin{equation}
j^{i}\partial _{i}\varphi =\frac{\partial (\varphi ,\tau _{1},\tau _{2},\tau
_{3})}{\partial (x^{0},x^{1},x^{2},x^{3})}  \label{a5.17}
\end{equation}

The action (\ref{a5.13a}) can be rewritten in the Lagrangian coordinates $%
\tau $ in the form 
\begin{equation}
\mathcal{A}_{\mathrm{Dqu}}[x,\mathbf{\xi }]=\int \left\{ -\kappa _{0}m\sqrt{%
\dot{x}^{i}\dot{x}_{i}}+\hbar {\frac{(\dot{\mathbf{\xi }}\times \mathbf{\xi }%
)\mathbf{z}}{2(1+\mathbf{\xi z})}}+\hbar \frac{(\dot{\mathbf{x}}\times \ddot{%
\mathbf{x}})\mathbf{\xi }}{2\sqrt{\dot{x}^{s}\dot{x}_{s}}(\sqrt{\dot{x}^{s}%
\dot{x}_{s}}+\dot{x}^{0})}\right\} d^{4}\tau  \label{a5.18}
\end{equation}
where the dot means the total derivative $\dot{x}^{s}\equiv dx^{s}/d\tau
_{0} $.\ $x=\left\{ x^{0},\mathbf{x}\right\} =\{x^{i}\}$, $\;i=0,1,2,3$, $%
\mathbf{\xi }=\{\xi ^{\alpha }\}$, $\alpha =1,2,3$ are considered to be
functions of the Lagrangian coordinates $\tau _{0}$, $\mathbf{\tau }=\{\tau
_{1},\tau _{2},\tau _{3}\}$. Here and in what follows the symbol $\times $
means the vector product of two 3-vectors. The quantity$\;\mathbf{z}$ is the
constant unit 3-vector (\ref{f1.13}). The term $j^{i}\partial _{i}\varphi $
is omitted, because it reduces to a Jacobian (\ref{a5.17}), which does not
contribute to dynamic equations. In fact, variables $x$ depend on $\mathbf{%
\tau }$ as on parameters, because the action (\ref{a5.18}) does not contain
derivatives with respect to $\tau _{\alpha }$, \ $\alpha =1,2,3$. Lagrangian
density of the action (\ref{a5.18}) does not contain independent variables $%
\tau $ explicitly. Hence, it may be written in the form 
\begin{equation}
\mathcal{A}_{\mathrm{Dqu}}[x,\mathbf{\xi }]=\int \mathcal{A}_{\mathrm{Dcl}%
}[x,\mathbf{\xi }]d\mathbf{\tau ,\qquad d\tau }=d\tau _{1}d\tau _{2}d\tau
_{3}  \label{b3.8}
\end{equation}
where 
\begin{equation}
\mathcal{S}_{\mathrm{Dcl}}:\qquad \mathcal{A}_{\mathrm{Dcl}}[x,\mathbf{\xi }%
]=\int \left\{ -\kappa _{0}m\sqrt{\dot{x}^{i}\dot{x}_{i}}+\hbar {\frac{(\dot{%
\mathbf{\xi }}\times \mathbf{\xi })\mathbf{z}}{2(1+\mathbf{\xi z})}}+\hbar 
\frac{(\dot{\mathbf{x}}\times \ddot{\mathbf{x}})\mathbf{\xi }}{2\sqrt{\dot{x}%
^{s}\dot{x}_{s}}(\sqrt{\dot{x}^{s}\dot{x}_{s}}+\dot{x}^{0})}\right\} d\tau
_{0}  \label{b3.9}
\end{equation}

The action (\ref{b3.8}) is the action for the dynamic system $\mathcal{S}_{%
\mathrm{Dqu}}$, which is a set of similar independent dynamic systems $%
\mathcal{S}_{\mathrm{Dcl}}$. Such a dynamic system is called a statistical
ensemble. Dynamic systems $\mathcal{S}_{\mathrm{Dcl}}$ are elements
(constituents) of the statistical ensemble $\mathcal{E}_{\mathrm{Dqu}}$.
Dynamic equations for each $\mathcal{S}_{\mathrm{Dcl}}$ form a system of
ordinary differential equations. It may be interpreted in the sense that the
dynamic system $\mathcal{S}_{\mathrm{Dcl}}$ may be considered to be a
classical one, although its Lagrangian contains the quantum constant $\hbar $%
. The dynamic system $\mathcal{S}_{\mathrm{Dcl}}$ will be referred to as the
classical Dirac particle.

Note that the quantum constant $\hbar $ can be eliminated from the action (%
\ref{b3.9}) by means of the change of variables 
\begin{equation}
\mathbf{\xi }\rightarrow \mathbf{\Xi }=\hbar \mathbf{\xi },\qquad \mathbf{z}%
\rightarrow \mathbf{Z}=\frac{1}{\hbar }\mathbf{z},\qquad \mathbf{\Xi }%
^{2}=\hbar ^{2},\qquad \mathbf{Z}^{2}=\frac{1}{\hbar ^{2}}  \label{b3.9a}
\end{equation}

The first term in the action (\ref{b3.9}) is relativistic. It describes a
motion of classical Dirac particle as a whole. The last two terms in the
action (\ref{b3.9}) are nonrelativistic. They describe some internal degrees
of freedom of the classical Dirac particle. This internal motion (classical
zitterbewegung) means that the classical Dirac particle has some internal
structure which is described by a method incompatible with relativity
principles. Maybe, the classical Dirac particle is composite. It should be
considered to be consisting of several pointlike particles. At any rate the
classical Dirac particle is not a pointlike particle. It has a more
complicated structure which is described by the variable $\mathbf{\xi }$ and
by the second order derivative $\mathbf{\ddot{x}}$.

It is easy to see that the action (\ref{b3.9}) is invariant with respect to
transformation $\tau _{0}\rightarrow \tilde{\tau}_{0}=F\left( \tau
_{0}\right) $, where $F$ is an arbitrary monotone function. This
transformation admits one to choose the variable $t=x^{0}$ as a parameter $%
\tau _{0},$ or to choose the parameter $\tau _{0}$ in such a way that $\dot{x%
}^{l}\dot{x}_{l}=\dot{x}_{0}^{2}-\mathbf{\dot{x}}^{2}=1$. In the last case
the parameter $\tau _{0}$ is the proper time along the world line of
classical Dirac particle. Besides, invariance with respect to transformation 
$\tau _{0}\rightarrow \tilde{\tau}_{0}=F\left( \tau _{0}\right) $ leads to a
connection between the components of the canonical momentum 
\[
p_{k}=\frac{\partial L}{\partial \dot{x}^{k}}-\frac{d}{d\tau _{0}}\frac{%
\partial L}{\partial \ddot{x}^{k}},\qquad k=0,1,2,3 
\]
where $L$ is the Lagrange function for the action (\ref{b3.9}).

\section{Solution of dynamic equations for $\mathcal{S}_{\mathrm{Dcl}}$}

We shall not consider here problems connected with relativistic
non-invariance of terms, describing internal degrees of freedom, referring
to \cite{R2001}, where these problems are discussed. We obtain dynamic
equations generated by the action (\ref{b3.9}), solve them and try to
interpret the obtained solution.

Variation of the action (\ref{b3.9}) with respect to $\mathbf{x}$ gives the
dynamic equation 
\begin{equation}
\frac{d}{d\tau _{0}}\left( -\kappa _{0}m\frac{\mathbf{\dot{x}}}{\sqrt{\dot{x}%
^{s}\dot{x}_{s}}}+\frac{\hbar Q}{2}(\mathbf{\xi }\times \ddot{\mathbf{x}})-%
\frac{\hbar }{2}\frac{\partial Q}{\partial \mathbf{\dot{x}}}(\dot{\mathbf{x}}%
\times \ddot{\mathbf{x}})\mathbf{\xi }+\frac{\hbar }{2}\frac{d}{d\tau _{0}}%
\left( Q(\mathbf{\xi }\times \mathbf{\dot{x}})\right) \right) =0
\label{b6.1}
\end{equation}
where 
\begin{equation}
Q=Q\left( \dot{x}\right) =\left( \sqrt{\dot{x}^{s}\dot{x}_{s}}(\sqrt{\dot{x}%
^{s}\dot{x}_{s}}+\dot{x}^{0})\right) ^{-1},\qquad \dot{x}^{s}\dot{x}_{s}=%
\dot{x}_{0}^{2}-\mathbf{\dot{x}}^{2}  \label{b6.2}
\end{equation}
Varying the action (\ref{b3.9}) with respect to $x^{0}$, we obtain 
\begin{equation}
\frac{d}{d\tau _{0}}\left( \kappa _{0}m\frac{\dot{x}^{0}}{\sqrt{\dot{x}^{s}%
\dot{x}_{s}}}-\frac{\hbar }{2}\frac{\partial Q}{\partial \dot{x}^{0}}(\dot{%
\mathbf{x}}\times \ddot{\mathbf{x}})\mathbf{\xi }\right) =0  \label{b6.1a}
\end{equation}

Varying the action (\ref{b3.9}) with respect to $\mathbf{\xi }$\textbf{,}
one should take into account the side constraint $\mathbf{\xi }^{2}=1$.
Setting 
\begin{equation}
\xi ^{\alpha }=\frac{\zeta ^{\alpha }}{\sqrt{\mathbf{\zeta }^{2}}},\qquad
\alpha =1,2,3  \label{b6.3}
\end{equation}
where $\mathbf{\zeta }$ is an arbitrary 3-pseudovector, one obtains 
\begin{equation}
\frac{\delta \mathcal{A}_{\mathrm{dcl}}}{\delta \zeta ^{\mu }}=\frac{\delta 
\mathcal{A}_{\mathrm{dcl}}}{\delta \xi ^{\alpha }}\frac{\delta \xi ^{\alpha }%
}{\delta \zeta ^{\mu }}=\frac{\delta \mathcal{A}_{\mathrm{dcl}}}{\delta \xi
^{\alpha }}\frac{\delta ^{\alpha \mu }-\xi ^{\alpha }\xi ^{\mu }}{\sqrt{%
\mathbf{\zeta }^{2}}}=0  \label{b6.4}
\end{equation}
It means that there are only two independent equations among three dynamic
equations (\ref{b6.4}). They are orthogonal to 3-pseudovector $\mathbf{\xi }$
and can be obtained from equation $\delta \mathcal{A}_{\mathrm{dcl}}/\delta
\xi ^{\alpha }=0$ by means of vector product with $\mathbf{\xi }$. 
\begin{equation}
-\hbar {\frac{\left( \dot{\mathbf{\xi }}\times \mathbf{z}\right) \times 
\mathbf{\xi }}{2(1+\mathbf{z\xi })}+}\hbar \left( {-}\frac{d}{d\tau _{0}}{{%
\frac{(\mathbf{\xi }\times \mathbf{z})}{2(1+\mathbf{z\xi })}}-\frac{(\dot{%
\mathbf{\xi }}\times \mathbf{\xi })\mathbf{z}}{2(1+\mathbf{z\xi })^{2}}%
\mathbf{z}}\right) \times \mathbf{\xi }+\hbar \frac{(\dot{\mathbf{x}}\times 
\ddot{\mathbf{x}})\times \mathbf{\xi }}{2}Q=0  \label{b6.5}
\end{equation}
After transformations this equation reduces to the equation (see Appendix\
B) 
\begin{equation}
\mathbf{\dot{\xi}}=-(\mathbf{\dot{x}}\times \mathbf{\ddot{x}})\times \mathbf{%
\xi }Q,  \label{c7.1}
\end{equation}
which does not contain the vector $\mathbf{z}$. It means that $\mathbf{z}$
determines a fictitious direction in the space-time. Note that $\mathbf{z}$
in the action (\ref{c4.15}) for the system $\mathcal{S}_{\mathrm{D}}$ is
fictitious also, because the term containing $\mathbf{z}$ is the same in
both actions (\ref{c4.15}) for $\mathcal{S}_{\mathrm{D}}$ and (\ref{a5.9})
for $\mathcal{S}_{\mathrm{Dqu}}$.

Using invariance of the action (\ref{b3.9}) with respect to transformation
of the parameter $\tau _{0}$, we choose $\tau _{0}$ in such a way, that 
\begin{equation}
\sqrt{\dot{x}^{s}\dot{x}_{s}}=\sqrt{\dot{x}_{0}^{2}-\mathbf{\dot{x}}^{2}}%
=1,\qquad \dot{x}_{0}=\sqrt{1+\mathbf{\dot{x}}^{2}}  \label{b6.10}
\end{equation}
Then, using condition (\ref{b6.10}), we obtain from (\ref{b6.2}) for
quantities $Q$, $\partial Q/\partial \dot{x}_{0}$, $\partial Q/\partial 
\mathbf{\dot{x}}$ 
\begin{equation}
Q=\frac{1}{1+\dot{x}_{0}},\qquad \frac{\partial Q}{\partial \dot{x}_{0}}%
=-1,\qquad \frac{\partial Q}{\partial \mathbf{\dot{x}}}=\frac{\mathbf{\dot{x}%
}\left( 2+\dot{x}_{0}\right) }{\left( 1+\dot{x}_{0}\right) ^{2}}
\label{b6.11}
\end{equation}

Integration of equation (\ref{b6.1a}) leads to

\begin{equation}
\kappa _{0}m\dot{x}_{0}+{\frac{\hbar }{2}}\left( \mathbf{\dot{x}}\times 
\mathbf{\ddot{x}}\right) \mathbf{\xi }=-p_{0}  \label{b6.7}
\end{equation}
where $p_{0}$ is an integration constant. This constant $p_{0}$ describes
the time component of the dynamic system $\mathcal{S}_{\mathrm{Dcl}}$
canonical 4-momentum.

Integration of equation (\ref{b6.1}) gives 
\begin{equation}
-\kappa _{0}m\frac{\mathbf{\dot{x}}}{\sqrt{\dot{x}^{s}\dot{x}_{s}}}+\frac{%
\hbar Q}{2}(\mathbf{\xi }\times \ddot{\mathbf{x}})-\frac{\hbar }{2}\frac{%
\partial Q}{\partial \mathbf{\dot{x}}}(\dot{\mathbf{x}}\times \ddot{\mathbf{x%
}})\mathbf{\xi }+\frac{\hbar }{2}\frac{d}{d\tau _{0}}\left( Q(\mathbf{\xi }%
\times \mathbf{\dot{x}})\right) =-\mathbf{p=}\text{const}  \label{e6.8}
\end{equation}
where $\mathbf{p}$ is the 3-momentum of the dynamic system $\mathcal{S}_{%
\mathrm{Dcl}}$ as a whole.

Using the gauge (\ref{b6.2}) and relations (\ref{b6.11}), we rewrite the
equation (\ref{e6.8}) in the form 
\begin{equation}
-m\mathbf{\dot{x}}+\frac{\hbar }{2}\frac{(\mathbf{\xi }\times \ddot{\mathbf{x%
}})}{1+\dot{x}_{0}}-\frac{\hbar }{2}\frac{\mathbf{\dot{x}}\left( 2+\dot{x}%
_{0}\right) }{\left( 1+\dot{x}_{0}\right) ^{2}}(\dot{\mathbf{x}}\times \ddot{%
\mathbf{x}})\mathbf{\xi }+\frac{\hbar }{2}\frac{d}{d\tau _{0}}\left( \frac{(%
\mathbf{\xi }\times \mathbf{\dot{x}})}{1+\dot{x}_{0}}\right) =-\mathbf{p}
\label{e6.8b}
\end{equation}
If we set $\hbar =0$ in (\ref{e6.8b}), we obtain conventional connection $%
\mathbf{p}=m\mathbf{\dot{x}}$ between the velocity $\mathbf{\dot{x}}=d%
\mathbf{x}/d\tau _{0}$ and the momentum of a free particle. But the quantum
constant $\hbar $ is a coefficient before the highest time derivative, and
setting $\hbar =0$, we suppress some degrees of freedom.

If these additional degrees of freedom are not excited (or suppressed), the
classical Dirac particle has six degrees of freedom. We shall see that
characteristic energy associated with additional degrees of freedom is of
the order of the particle rest energy $m$. At low energetic processes
(calculation of atomic spectra, quantum electrodynamics) one may neglect
these degrees of freedom, remaining only numerical characteristics (spin,
magnetic momentum) of these degrees of freedom. However, in the case of high
energies (ultrarelativistic collisions, structure of elementary particles),
one cannot neglect these degrees of freedom. Of course, using the Dirac
equation, we take into account these additional degrees of freedom
automatically. But it is important also to take into account these
additional degrees of freedom in our interpretation of the high energetic
processes.

Transformation and solution of equation (\ref{e6.8}) is rather bulky. Many
efforts is used to prove that the 3-vectors $\mathbf{\xi }$, $\mathbf{\dot{x}%
}$, and $\mathbf{\ddot{x}}$ are mutually orthogonal and their modules are
constant \cite{R2001} in the coordinate system, where $\mathbf{p}=0$. We
shall not spend time for this proof. Instead, we choose the coordinate
system in such a way that $\mathbf{p}=0$ 
\begin{equation}
\mathbf{\xi =}\left\{ 0,0,\varepsilon _{0}\right\} ,\qquad \varepsilon
_{0}=\pm 1  \label{e6.8a}
\end{equation}
and impose constraints 
\begin{equation}
\mathbf{\dot{x}}^{2}=\text{const,\qquad }\left( \mathbf{\dot{x}\xi }\right)
=0,\qquad \left( \mathbf{\ddot{x}\xi }\right) =0,\qquad \left( \mathbf{\dot{x%
}}\times \mathbf{\ddot{x}}\right) \mathbf{\xi }=\text{const}  \label{e6.9}
\end{equation}
We use constraints (\ref{e6.9}) in solution of the system of dynamic
equations (\ref{c7.1}), (\ref{b6.7}), (\ref{e6.8b}) and show that the
constraints (\ref{e6.9}) are compatible with dynamic equations (\ref{c7.1}),
(\ref{b6.7}), (\ref{e6.8b}).

Taking into account (\ref{e6.9}) and (\ref{b6.10}), we introduce new
variables 
\begin{equation}
\mathbf{y=}\frac{\mathbf{\dot{x}}}{\sqrt{1+\dot{x}_{0}}}\mathbf{=}\frac{%
\mathbf{\dot{x}}}{\sqrt{1+\sqrt{1+\mathbf{\dot{x}}^{2}}}},\;\;\;\;\;\;%
\mathbf{\dot{x}=y}\sqrt{\left( \mathbf{y}^{2}+2\right) }  \label{b6.13}
\end{equation}
\begin{equation}
\dot{x}_{0}=\sqrt{1+\mathbf{y}^{2}\left( \mathbf{y}^{2}+2\right) }=\mathbf{y}%
^{2}+1  \label{b6.14}
\end{equation}
Introducing designation 
\begin{equation}
\mathbf{y}^{2}=\gamma -1=\text{const,}  \label{b6.14b}
\end{equation}
we obtain 
\begin{equation}
\dot{x}_{0}=\sqrt{1+\mathbf{y}^{2}\left( \mathbf{y}^{2}+2\right) }=\mathbf{y}%
^{2}+1=\gamma =\text{const}  \label{b6.14a}
\end{equation}
Then at $\mathbf{p}=0$ the equation (\ref{e6.8b}) takes the form 
\begin{equation}
-\kappa _{0}m\mathbf{y}\left( \gamma +1\right) +\frac{\hbar }{2}(\mathbf{\xi 
}\times \mathbf{\dot{y}})-\frac{\hbar }{2}\left( \gamma +2\right) \left( (%
\mathbf{y}\times \mathbf{\dot{y}})\mathbf{\xi }\right) \mathbf{y}+\frac{%
\hbar }{2}\frac{d}{d\tau _{0}}\left( (\mathbf{\xi }\times \mathbf{y})\right)
=0  \label{e6.15}
\end{equation}
The equation (\ref{c7.1}) takes the form 
\begin{equation}
\mathbf{\dot{\xi}}=-(\mathbf{y}\times \mathbf{\dot{y}})\times \mathbf{\xi }=0
\label{e6.17}
\end{equation}
because of constraints (\ref{e6.9}). In terms of variables $\mathbf{y}$
conditions (\ref{e6.9}) have the form 
\begin{equation}
\mathbf{y}^{2}=\gamma -1,\qquad \left( \mathbf{\xi y}\right) =0,\qquad
\left( \mathbf{\xi \dot{y}}\right) =0,\qquad \left( \mathbf{y\dot{y}}\right)
=0  \label{e6.18}
\end{equation}
where $\gamma $ is a constant of integration\textbf{. }In accordance with (%
\ref{b6.14a}) and (\ref{e6.18}) we obtain 
\begin{equation}
(\mathbf{y}\times \mathbf{\dot{y}})\mathbf{\xi }=\varepsilon _{0}\omega
\left( \gamma -1\right)  \label{e6.19}
\end{equation}
where $\omega $ is an indefinite constant (some angular velocity).

Substituting (\ref{e6.19}) in (\ref{e6.15}), we obtain after simplification 
\begin{equation}
(\mathbf{\xi }\times \mathbf{\dot{y}})-\left( \frac{1}{2}\left( \gamma
+2\right) \left( \gamma -1\right) \varepsilon _{0}\omega +\frac{\kappa _{0}m%
}{\hbar }\left( \gamma +1\right) \right) \mathbf{y}=0  \label{e6.20}
\end{equation}
As far as $\mathbf{y}^{2}=\gamma -1$, the equation (\ref{e6.19}) describes
rotation of the vector $\mathbf{y}$ with the angular frequency $\omega $.
Equation (\ref{e6.20}) describes rotation of the vector $\mathbf{y}$ around
the vector $\mathbf{\xi }$ with the angular frequency $\frac{1}{2}\left(
\gamma +2\right) \left( \gamma -1\right) \varepsilon _{0}\omega |\frac{%
\kappa _{0}m}{\hbar }\left( \gamma +1\right) $. Equations (\ref{e6.19}) and (%
\ref{e6.20}) are compatible, if these frequencies coincide. According to (%
\ref{e6.18}) vectors $\mathbf{y}$ and $\mathbf{\dot{y}}$ are orthogonal to $%
\mathbf{\xi }$\textbf{.} Then in accordance with (\ref{e6.8a}) the vectors $%
\mathbf{y}$ and $\mathbf{\dot{y}}$ can be represented in the form 
\begin{eqnarray}
\mathbf{y} &=&\left\{ \sqrt{\gamma -1}\cos \Phi ,\sqrt{\gamma -1}\sin \Phi
,0\right\}  \label{e6.21} \\
\mathbf{\dot{y}} &=&\left\{ -\sqrt{\gamma -1}\omega \sin \Phi ,\sqrt{\gamma
-1}\omega \cos \Phi ,0\right\} ,\qquad \omega =\frac{d\Phi }{d\tau _{0}}
\label{e6.22}
\end{eqnarray}
By means of (\ref{e6.21}), and (\ref{e6.22}) the equations (\ref{e6.20})
take the form 
\begin{eqnarray}
-\varepsilon _{0}\omega y_{1}-\left( \frac{1}{2}\left( \gamma +2\right)
\left( \gamma -1\right) \varepsilon _{0}\omega +\frac{\kappa _{0}m}{\hbar }%
\left( \gamma +1\right) \right) y_{1} &=&0  \label{e6.23} \\
-\varepsilon _{0}\omega y_{2}-\left( \frac{1}{2}\left( \gamma +2\right)
\left( \gamma -1\right) \varepsilon _{0}\omega +\frac{\kappa _{0}m}{\hbar }%
\left( \gamma +1\right) \right) y_{2} &=&0  \label{e6.24}
\end{eqnarray}
Equations (\ref{e6.23}), (\ref{e6.24}) are satisfied, provided 
\begin{equation}
\varepsilon _{0}\omega +\left( \frac{1}{2}\left( \gamma +2\right) \left(
\gamma -1\right) \varepsilon _{0}\omega +\frac{\kappa _{0}m}{\hbar }\left(
\gamma +1\right) \right) =0  \label{e6.25}
\end{equation}

Solution of (\ref{e6.25}) has the form 
\begin{equation}
\omega =-\frac{2\varepsilon _{0}\kappa _{0}m}{\hbar \gamma }  \label{e6.26}
\end{equation}
According to (\ref{b6.13}) and (\ref{b6.14}) the dynamic equation (\ref{b6.7}%
) takes the form 
\begin{equation}
-p_{0}=\kappa _{0}m\gamma +{\frac{\hbar }{2}}\left( \mathbf{y}\times \mathbf{%
\dot{y}}\right) \mathbf{\xi }\left( \gamma +1\right)   \label{e6.27}
\end{equation}
Using relations (\ref{e6.19}) and (\ref{e6.26}) we obtain from (\ref{e6.27}) 
\begin{equation}
-p_{0}=\kappa _{0}m\left( \gamma -\frac{\gamma ^{2}-1}{\gamma }\right) =%
\frac{\kappa _{0}m}{\gamma },\qquad \kappa _{0}=\pm 1  \label{e6.27a}
\end{equation}
Then we obtain for the rest mass $M$ of the dynamic system $\mathcal{S}_{%
\mathrm{Dcl}}$. 
\begin{equation}
M_{\mathrm{Dcl}}=\sqrt{p_{0}^{2}-\mathbf{p}^{2}}=\left| p_{0}\right| =\frac{m%
}{\gamma }  \label{e6.28}
\end{equation}

Note, that writing the relation (\ref{e6.28}), we do not act quite
consequently. Writing the relation (\ref{e6.28}), we suppose that the
dynamic equations (\ref{b6.7}) and (\ref{e6.8}) are relativistically
invariant, and solution of equations (\ref{b6.7}), (\ref{e6.8}) for
arbitrary $\mathbf{p}$ can be obtained from the solution for $\mathbf{p}=0$
by means of a corresponding Lorentz transformation. Unfortunately, dynamic
equations (\ref{b6.7}), (\ref{e6.8}) are not relativistically invariant, and
for arbitrary $\mathbf{p}$ the solution is not a helix, in general, although
it is a helix for $\mathbf{p}=0$. World line is a helix approximately in the
nonrelativistic case, when $\left| \mathbf{p}\right| \ll m$.

Let us transit from independent variable $\tau _{0}$ to the independent
variable $x^{0}=t$. We have 
\begin{equation}
\Omega t=-\varepsilon _{0}\kappa _{0}\omega \tau _{0},\qquad -\varepsilon
_{0}\kappa _{0}\omega =\Omega \dot{x}_{0}=\Omega \gamma =\frac{2m}{\hbar
\gamma },\qquad \Omega =\frac{2m}{\hbar \gamma ^{2}}  \label{e6.29}
\end{equation}
Returning from variables $\mathbf{y}$ to variables $\mathbf{\dot{x}}$\textbf{%
, }we obtain instead of (\ref{e6.21}) and (\ref{e6.22}) 
\begin{eqnarray}
\frac{d\mathbf{x}}{dt} &=&\left\{ \frac{\sqrt{\gamma ^{2}-1}}{\gamma }\cos
\left( \Omega t\right) ,-\frac{\sqrt{\gamma ^{2}-1}}{\gamma }\sin \left(
\Omega t\right) ,0\right\} ,\qquad \Omega =\frac{2m}{\hbar \gamma ^{2}}
\label{e6.30} \\
\mathbf{x} &=&\left\{ \frac{\hbar \gamma \sqrt{\gamma ^{2}-1}}{2m}\sin
\left( \frac{2m}{\hbar \gamma ^{2}}t\right) ,\frac{\hbar \gamma \sqrt{\gamma
^{2}-1}}{2m}\cos \left( \frac{2m}{\hbar \gamma ^{2}}t\right) ,0\right\}
\label{e6.31}
\end{eqnarray}
where $\gamma \geq 1$ is an arbitrary constant.

Thus, in the coordinate system, where the canonical momentum four-vector has
the form 
\begin{equation}
P_{k}=\left\{ p_{0},\mathbf{p}\right\} =\left\{ -\frac{\kappa _{0}m}{\gamma }%
,0,0,0\right\}  \label{e6.31a}
\end{equation}
the world line of the classical Dirac particle is a helix, which is
described by the relation 
\begin{eqnarray}
\left\{ t,\mathbf{x}\right\} &=&\left\{ t,a_{\mathrm{Dcl}}\sin \left( \Omega
t\right) ,a_{\mathrm{Dcl}}\cos \left( \omega _{\mathrm{Dcl}}t\right)
,0\right\}  \label{e6.33} \\
a_{\mathrm{Dcl}} &=&\frac{\hbar \gamma \sqrt{\gamma ^{2}-1}}{2m},\qquad
\omega _{\mathrm{Dcl}}=\frac{2m}{\hbar \gamma ^{2}}  \label{e6.34}
\end{eqnarray}

It follows from (\ref{e6.30}) that the classical Dirac particle velocity $%
\mathbf{v}=d\mathbf{x}/dt$ is expressed as follows 
\begin{equation}
\mathbf{v}^{2}=1-\frac{1}{\gamma ^{2}},\qquad \gamma =\frac{1}{\sqrt{1-%
\mathbf{v}^{2}}}  \label{e6.32}
\end{equation}
In other words, the quantity $\gamma $ is the Lorentz factor of the
classical Dirac particle.

We see that the characteristic frequency, connected with the internal
degrees of freedom is $2m/\gamma ^{2}$, and the characteristic energy is of
the order $\left| -m\gamma +m\gamma ^{-1}\right| $.

Parameters $\gamma $ and $\omega _{\mathrm{Dcl}}$ as functions of the radius 
$a_{\mathrm{Dcl}}$ and the Dirac mass $m$ have the form 
\begin{equation}
\gamma =\sqrt{\frac{1}{2}\left( 1+\sqrt{1+\zeta ^{2}}\right) },\qquad \omega
_{\mathrm{Dcl}}=\frac{4m}{\hbar \left( 1+\sqrt{1+\zeta ^{2}}\right) }\qquad
\zeta =\frac{4ma_{\mathrm{Dcl}}}{\hbar }  \label{e6.35}
\end{equation}

\section{Dynamical interpretation of the classical Dirac particle}

If coordinates $x^{k}=\left\{ t,\mathbf{x}\right\} $ are interpreted as
coordinates of classical Dirac particle, it seems rather strange, that the
world line of a free particle is a helix, but not the straight line. Why
does the free classical particle rotate in the coordinate system, where
total momentum $\mathbf{p}=0$? Note, that the coordinates of the free
quantum Dirac particle $\mathcal{S}_{\mathrm{D}}$ contain oscillating
component, whereas momentum of $\mathcal{S}_{\mathrm{D}}$ does not contain
oscillating component \cite{D58} (sec. 69). This oscillating motion with the
frequency $\omega \geq 2m/\hbar $ is known as zitterbewegung. Usually the
zitterbewegung is considered to be a specific quantum phenomenon, but here
we obtain classical analog of the zitterbewegung and this classical
description contains the quantum constant $\hbar $.

Classical dynamic system $\mathcal{S}_{\mathrm{Dcl}}$ contains four
rotational degrees of freedom in addition to six conventional translation
degrees of freedom. It means that the classical dynamic system $\mathcal{S}_{%
\mathrm{Dcl}}$ is not a pointlike particle, because it has internal degrees
of freedom. How does one interpret these additional degrees of freedom? It
seems that the dynamic system $\mathcal{S}_{\mathrm{Dcl}}$ consists of
several constituents, rotating around its center of inertia. This idea is
found in accordance with the contemporary ideas, that such Dirac particles
as the proton and the neutron consist of quarks.

We investigate, to what extent the classical dynamic system $\mathcal{S}_{%
\mathrm{Dcl}}$ may be interpreted is a rotator. Rotator is a dynamic system $%
\mathcal{S}_{\mathrm{r}},$ consisting of two coupled particles of mass $%
m_{0} $, which can rotate around their center of mass. The distance between
particles is to be constant, i.e. the particles are not to vibrate. Rigid
nonrelativistic rotator $\mathcal{S}_{\mathrm{nr}}$ is described by the
action 
\begin{equation}
\mathcal{S}_{\mathrm{nr}}:\qquad \mathcal{A}\left[ \mathbf{x}_{1},\mathbf{x}%
_{2},\mu \right] =\int \left( \sum\limits_{k=1}^{k=2}\frac{m_{0}\mathbf{\dot{%
x}}_{k}^{2}}{2}+\mu \left( \left( \mathbf{x}_{1}-\mathbf{x}_{2}\right)
^{2}-4a^{2}\right) \right) dt  \label{b8.46a}
\end{equation}
where $2a$ is the distance (length of string) between the particles. The
parameter $a$ is determined by the length of rigid coupling between two
particles. It does not depend on initial conditions. The angular frequency $%
\omega $ of rotation is an arbitrary constant of integration, which does not
connect with the length $2a$ of the string.

If the coupling is elastic, the action for $\mathcal{S}_{\mathrm{nr}}$
should be written in the form 
\begin{equation}
\mathcal{S}_{\mathrm{nr}}:\qquad \mathcal{A}\left[ \mathbf{x}_{1},\mathbf{x}%
_{2},\mu \right] =\int \left( \sum\limits_{k=1}^{k=2}\frac{m_{0}\mathbf{\dot{%
x}}_{k}^{2}}{2}-U\left( \left| \mathbf{x}_{1}-\mathbf{x}_{2}\right| \right)
\right) dt  \label{b8.46b}
\end{equation}
where $U$ is the potential energy, describing interaction energy between two
particles. The dynamic equations for relative motion of the particles have
the form 
\begin{equation}
-m_{0}\left( \mathbf{\ddot{x}}_{1}-\mathbf{\ddot{x}}_{2}\right) -2\frac{%
\left( \mathbf{x}_{1}-\mathbf{x}_{2}\right) }{\left| \mathbf{x}_{1}-\mathbf{x%
}_{2}\right| }U^{\prime }\left( \left| \mathbf{x}_{1}-\mathbf{x}_{2}\right|
\right) =0,\qquad U^{\prime }\left( \left| \mathbf{x}_{1}-\mathbf{x}%
_{2}\right| \right) =\frac{\partial U\left( \left| \mathbf{x}_{1}-\mathbf{x}%
_{2}\right| \right) }{\partial \left| \mathbf{x}_{1}-\mathbf{x}_{2}\right| }
\label{b8.47d}
\end{equation}
Equations (\ref{b8.47d}) describe both rotation and radial vibrations of the
particle. Let us imagine that for some reason the vibrations are dumped, and
only rotational motion retains. It means that 
\begin{equation}
\left| \mathbf{x}_{1}-\mathbf{x}_{2}\right| =2a=\text{const},\qquad
U^{\prime }\left( \left| \mathbf{x}_{1}-\mathbf{x}_{2}\right| \right) =\text{%
const}  \label{b8.47e}
\end{equation}
and dynamic equations (\ref{b8.47d}) turns into linear equations for
variables $\mathbf{x}_{1}-\mathbf{x}_{2}$. Solution of these equations
describes a rotation in any 2-plane. For instance, in the plane $XY$ we
obtain 
\begin{equation}
\mathbf{x}_{1}-\mathbf{x}_{2}=\left\{ 2a\cos \left( \omega t\right) ,2a\sin
\left( \omega t\right) ,0\right\}  \label{b8.47f}
\end{equation}
where the angular frequency 
\begin{equation}
\omega =\omega \left( a\right) =\sqrt{\frac{U^{\prime }\left( 2a\right) }{%
m_{0}a}}  \label{b8.47g}
\end{equation}
is not an arbitrary constant. It depends on the form of the potential energy
of the elastic string and of radius $a$.

Unfortunately, in the relativistic case one cannot introduce potential
energy of interaction between two particles. But we may introduce the
rigidity function $f_{\mathrm{r}}\left( a\right) $, defining it by the
relation 
\begin{equation}
f_{\mathrm{r}}\left( a\right) =\frac{M-2m_{0}}{2m_{0}}  \label{b8.48h}
\end{equation}
where $M$ is the total mass of the rotator at the state of rotation, and $%
2m_{0}$ is its mass at rest. In the nonrelativistic case 
\[
M=2m_{0}+2m_{0}\frac{v^{2}}{2c^{2}}=2m_{0}\left( 1+\frac{a^{2}\omega ^{2}}{%
2c^{2}}\right) =2m_{0}\left( 1+\frac{a^{2}}{m_{0}c^{2}}U^{\prime }\left(
2a\right) \right) 
\]
\begin{equation}
f_{\mathrm{r}}\left( a\right) =\frac{a}{m_{0}c^{2}}U^{\prime }\left(
2a\right)  \label{b8.48j}
\end{equation}
The relation (\ref{b8.48j}) connects the rigidity function with the
potential energy of the string in the nonrelativistic case. This relation
may be extrapolated to the relativistic case, when the potential energy of
the string cannot be introduced. The rigidity function $f_{\mathrm{r}}\left(
a\right) $ of the established rotation of the relativistic rotator may be
introduced by means of the relation (\ref{b8.48h}). Then by means of the
relation (\ref{b8.48j}) one can introduce formally the potential energy of
interaction between the constituents of the relativistic rotator. We may
also describe interaction between the constituents directly in terms of the
rigidity function.

Solution for the dynamic system (\ref{b8.46a}) can be reduced to
''established'' solution for the dynamic system (\ref{b8.46b}), if we
introduce a proper coupling between the constants of integration $a$ and $%
\omega $, or introduce a proper rigidity function.

Solution (\ref{e6.31}) for the classical Dirac particle $\mathcal{S}_{%
\mathrm{Dcl}}$ is a helix. It reminds the solution for a rotator, but the
identification is rather complicated, because instead of mass $m_{0}$ of the
rotator constituents, we have the Dirac mass $m$, whose meaning is unclear.
To produce such an identification, we need to solve dynamic equations for
the relativistic rotator of the type (\ref{b8.46a}). Comparing the obtained
solution with (\ref{e6.31}), we determine the relation between the masses $m$
and $m_{0}$. Thereafter we can determine the rigidity function and evaluate
the character of interaction between the classical Dirac particle
constituents.

In the relativity theory a rigid coupling is impossible. Also there are no
reasons for introduction of a potential energy of interaction between the
particles, because in the relativity theory a long-range action is absent.
We are forced to choose another way.

Let us consider established relativistic motion of two particles of mass $%
m_{0}$, coupled between themselves by a massless elastic string. The
condition of established motion means that the particles move in such a way
that the length of the string does not change, and one may neglect degrees
of freedom, connected with the string. Mathematically it means, that there
is such a coordinate system $K$ (maybe, rotating), where particles are at
rest.

Let $\mathcal{L}_{1}$ and $\mathcal{L}_{2},$ be world lines of particles 
\begin{equation}
\mathcal{L}_{k}:\qquad x_{(k)}^{i}=x_{(k)}^{i}\left( \tau _{k}\right)
,\qquad i=0,1,2,3;\;\;\;k=1,2  \label{b8.1}
\end{equation}
where $\tau _{k},\;\;k=1,2$ are parameters along these world lines. From
geometrical viewpoint the steady (established) motion of particles means
that any spacelike 3-plane $\mathcal{S}$, crossing $\mathcal{L}_{1}$
orthogonally, crosses $\mathcal{L}_{2}$ also orthogonally. This circumstance
permits one to synchronize events on $\mathcal{L}_{1}$ and on $\mathcal{L}%
_{2}$.

Let parameters $\tau _{1}$ and $\tau _{2}$ be chosen in such a way that they
have the same value $\tau $ at points $\mathcal{L}_{1}\cap \mathcal{S}$ and $%
\mathcal{L}_{2}\cap \mathcal{S}$. The steady state conditions are written in
the form 
\begin{equation}
\frac{dx_{(1)}^{i}\left( \tau \right) }{d\tau }\left( x_{(1)i}\left( \tau
\right) -x_{(2)i}\left( \tau \right) \right) =\frac{dx_{(2)}^{i}\left( \tau
\right) }{d\tau }\left( x_{(1)i}\left( \tau \right) -x_{(2)i}\left( \tau
\right) \right) =0  \label{b8.2}
\end{equation}

Let us describe motion of the two particles (constituents) by the action 
\begin{equation}
\mathcal{S}_{\mathrm{rr}}:\qquad \mathcal{A}_{\mathrm{rr}}\left[ x_{\left(
1\right) },x_{\left( 2\right) }\right] =-\int \sum\limits_{k=1}^{k=2}m_{0}%
\sqrt{\dot{x}_{\left( k\right) }^{l}\dot{x}_{\left( k\right) l}}d\tau
\label{b8.3}
\end{equation}
where variables $x_{\left( k\right) }=\left\{ x_{\left( k\right)
}^{0},x_{\left( k\right) }^{1},x_{\left( k\right) }^{2},x_{\left( k\right)
}^{3},\right\} $, $k=1,2$ are considered to be functions of the same
parameter $\tau $, $m_{0}$ is the mass of the rotator constituent, and the
period denotes differentiation with respect to $\tau $. World lines of
constituents are determined as extremals of the functional (\ref{b8.3}) with
side constraints (\ref{b8.2}).

We omit details of solution of the dynamic equations for the rotator (\ref
{b8.3}), (\ref{b8.2}) (One can find them in sec. 8 of \cite{R2001}). In the
coordinate system, where the rotator is at rest, the solution has the form 
\begin{eqnarray}
x_{(1)}^{i} &=&\left\{ t,\;a\cos \left( \omega _{0}t+\phi \right) ,\;a\sin
\left( \omega _{0}t+\phi \right) ,\;0\right\}  \label{b8.44} \\
x_{(2)}^{i} &=&\left\{ t,\;-a\cos \left( \omega _{0}t+\phi \right) ,\;-a\sin
\left( \omega _{0}t+\phi \right) ,\;0\right\}  \label{b8.45}
\end{eqnarray}
Here the quantities $a$, $\phi $, $\omega _{0}$ are constants of
integration. The angular frequency is connected with the total mass $M$ of
the rotator, and its rest mass $2m_{0}$ by means of relations 
\begin{equation}
M=\frac{2m_{0}}{\sqrt{1-a^{2}\omega _{0}^{2}}},\qquad \omega _{0}=\frac{%
\sqrt{M^{2}-4m_{0}^{2}}}{aM}  \label{f8.1}
\end{equation}
where $v=a\omega _{0}$ is the velocity of the rotator constituents.

One can see that both world lines $\mathcal{L}_{1}$ and $\mathcal{L}_{2}$
are helixes in the space-time. They describe rotation of two particles
around their common center of mass along a circle of radius $a$. Such a
dynamic system $\mathcal{S}_{\mathrm{rr}}$ may be qualified as a
relativistic rotator. Here $\omega _{0}$ is an arbitrary quantity, and
rotator (\ref{b8.3}), (\ref{b8.2}) is the dynamic system of type of (\ref
{b8.46a}), but not of (\ref{b8.46b}). The state of the rotator may be
described by the relative mass increase 
\begin{equation}
\alpha =\frac{\left( M-2m_{0}\right) }{2m_{0}}=\frac{1}{\sqrt{1-v^{2}}}-1=%
\frac{v^{2}}{\sqrt{1-v^{2}}\left( \sqrt{1-v^{2}}+1\right) }  \label{b8.76}
\end{equation}
which turns to the rigidity function (\ref{b8.48h}), provided we determine $%
\omega _{0}$ as a function of the radius $a$. Here $v=a\omega _{0}$ is the
velocity of a constituent in the coordinate system, where center of mass is
at rest. The quantity $\alpha $ is a part of total mass, conditioned by
rotation.

The radius $a$ and angular frequency $\omega _{0}$ of rotation are
independent constants of integration in the solution (\ref{b8.44}), (\ref
{b8.45}). For any real rotator the frequency $\omega _{0}$ and the radius $a$
cannot be independent. We suppose that the dynamic system $\mathcal{S}_{%
\mathrm{Dcl}}$, where angular velocity is coupled with the radius of helix,
is a special case of relativistic rotator $\mathcal{S}_{\mathrm{rr}}$. To
determine the rigidity function and the relation between $m$ and $m_{0}$, we
compare relations (\ref{f8.1}), (\ref{b8.44}) with relations (\ref{e6.34}), (%
\ref{e6.33}) and identify the quantities $\omega _{0}$, $M$, $a$ of dynamic
system $\mathcal{S}_{\mathrm{rr}}$ respectively with $\omega _{\mathrm{Dcl}}$%
, $M_{\mathrm{Dcl}}$, $a_{\mathrm{Dcl}}$ of dynamic system $\mathcal{S}_{%
\mathrm{Dcl}}$. Using relation (\ref{e6.35}) for $\gamma $ we obtain 
\begin{eqnarray}
\omega _{0} &=&\frac{\sqrt{M^{2}-4m_{0}^{2}}}{aM}=\frac{4m}{\hbar }\frac{1}{%
\left( \sqrt{1+\zeta ^{2}}+1\right) }=\omega _{\mathrm{Dcl}},\qquad M=\frac{m%
\sqrt{2}}{\sqrt{\left( \sqrt{1+\zeta ^{2}}+1\right) }}  \label{b8.71} \\
a &=&a_{\mathrm{Dcl}},\qquad \zeta =4a_{\mathrm{Dcl}}\frac{m}{\hbar },\qquad
c=1  \label{b8.72}
\end{eqnarray}

Resolving relations (\ref{b8.71}) with respect to variables $M$ and $m_{0}$,
one obtains all quantities of the relativistic rotator $\mathcal{S}_{\mathrm{%
rr}}$ in terms of parameters $\zeta =4ma_{\mathrm{Dcl}}/\hbar $ and $m$ of
the classical Dirac particle $\mathcal{S}_{\mathrm{Dcl}}$%
\begin{equation}
M=M_{\mathrm{Dcl}}=\frac{m\sqrt{2}}{\sqrt{\left( \sqrt{1+\zeta ^{2}}%
+1\right) }},\qquad m_{0}=\frac{m}{\left( \sqrt{1+\zeta ^{2}}+1\right) }
\label{b8.73}
\end{equation}
It follows from the second relation (\ref{b8.73}), that the relation between
the Dirac mass $m$ and the constituent mass $m_{0}$ depends on the state of
rotation. In other words, different states of the classical Dirac particle
correspond, in general, to different mass $m_{0}$ of the rotator
constituent. In this connection the question arises. What mass is principal $%
m$ or $m_{0}$?

One can express parameters $m,$ $M_{\mathrm{Dcl}},$ $\omega _{\mathrm{Dcl}},$
$v_{\mathrm{Dcl}}$ of $\mathcal{S}_{\mathrm{Dcl}}$ in terms of parameters $%
v=\zeta _{0}=4am_{0}/\hbar $, $m_{0}$ of the relativistic rotator $\mathcal{S%
}_{\mathrm{rr}}$. From the second relation (\ref{b8.73}) we obtain relation
between $\zeta $ and $\zeta _{0}$, 
\begin{equation}
\zeta _{0}=\frac{\zeta }{\sqrt{1+\zeta ^{2}}+1},\qquad \zeta =\frac{2\zeta
_{0}}{1-\zeta _{0}^{2}}  \label{b8.73a}
\end{equation}
After some calculations one obtains 
\begin{equation}
m=2m_{0}\frac{1}{1-v^{2}},\qquad M_{\mathrm{Dcl}}=\frac{2m_{0}}{\sqrt{1-v^{2}%
}},\qquad \omega _{\mathrm{Dcl}}=\frac{4m_{0}}{\hbar },  \label{b8.74}
\end{equation}
where 
\begin{equation}
v=\zeta _{0}=4a\frac{m_{0}}{\hbar }  \label{b8.75}
\end{equation}

It follows from the relations (\ref{b8.74}), (\ref{b8.75}) and (\ref{b8.48h}%
), that the rigidity function $f_{\mathrm{rDcl}}\left( a\right) $ for the
classical Dirac particle has the form 
\begin{equation}
f_{\mathrm{rDcl}}\left( a\right) =\frac{M_{\mathrm{Dcl}}-2m_{0}}{2m_{0}}=%
\frac{\hbar ^{2}}{\sqrt{\hbar ^{2}-\left( 4am_{0}\right) ^{2}}}-1
\label{e7.6}
\end{equation}

As far as two masses $m$ and $M_{\mathrm{Dcl}}$ are distinguished, the
question appears, which of the two masses is observable. To solve this
question, one needs to consider the Dirac dynamic system in the given
electromagnetic field.

We shall not discuss this question here. Instead, we discuss another
important problem. What force field does connect constituents of the Dirac
particle? Is it possible to separate these constituents as single particles?
The direct interaction between the remote constituents seems to be
incompatible with the relativity principles. It means that we are to
introduce gluons or some other bearers of interaction between the
constituents. But even in this case it is rather difficult to understand,
why the distance between the constituents is fixed at any state of the
dynamic system $\mathcal{S}_{\mathrm{Dcl}}$.

We shall see that the only solution of this problem is the geometrical
interpretation instead of the dynamical one.

\section{Geometrical model of the classical Dirac \newline
particle}

We explain what is the geometrical interpretation in the example of a
classical pointlike particle. Its motion is described by its world line $%
\mathcal{L}$%
\begin{equation}
\mathcal{L}:\qquad x^{i}=x^{i}\left( \tau \right) ,\qquad i=0,1,2,3
\label{e8.1}
\end{equation}
where $\tau $ is a real parameter along the world line. According to the
principles of relativity the world line (WL) is a physical object, whereas
the particle (pointlike object in 3-space) is an attribute of WL. We shall
use abbreviation WL (instead of world line), when we want to stress that the
world line is a physical object, but not an attribute of a particle (its
history). WL is not a completely geometrical object, because it contains a
non-geometric parameter: mass $m$. To make WL to be a geometrical object we
are to geometrize the mass. We make this as follows.

Instead of WL (\ref{e8.1}) we consider the broken line $\mathcal{T}_{\mathrm{%
br}}$%
\begin{equation}
\mathcal{T}_{\mathrm{br}}=\bigcup\limits_{i}\mathcal{T}_{\left[ P_{i}P_{i+1}%
\right] }  \label{e8.2}
\end{equation}
consisting of straight line segments $\mathcal{T}_{\left[ P_{i}P_{i+1}\right]
}$ of the same length $\mu $ 
\begin{equation}
\mathcal{T}_{\left[ P_{i}P_{i+1}\right] }=\left\{ R|\sqrt{2\sigma \left(
P_{i},P_{i+1}\right) }=\sqrt{2\sigma \left( P_{i},R\right) }+\sqrt{2\sigma
\left( R,P_{i+1}\right) }\right\}  \label{e8.3}
\end{equation}
where $\sigma $ is the world function of the space-time. Here$\ $the
space-time points $P_{i}$ , $i=0,\pm 1,\pm 2...$ are the break points of $%
\mathcal{T}_{\mathrm{br}}$. In the case, when the space-time is the
Minkowski space, the world function, written in the inertial coordinate
system, has the form 
\begin{equation}
\sigma =\sigma _{\mathrm{M}}\left( x,x^{\prime }\right) =\frac{1}{2}%
g_{ik}\left( x^{i}-x^{\prime i}\right) \left( x^{k}-x^{\prime k}\right)
\label{e8.4}
\end{equation}

In our geometrical description we use the fact that any physical geometry
can be described completely in terms of the world function $\sigma $ \cite
{R2002}. This circumstance allows one to use coordinateless description,
when all geometric objects and all relations between them are expressed in
terms of the world function $\sigma $. This method of description is
referred to as the $\sigma $-immanent description. It is convenient in the
sense, that a transition from one geometry to another one is carried out by
means of a change of the world function $\sigma $.

The vector $\mathbf{P}_{i}\mathbf{P}_{i+1}\equiv \overrightarrow{P_{i}P_{i+1}%
}=\left\{ P_{i},P_{i+1}\right\} $ is the ordered set of two points. It
describes the particle momentum on the segment $\mathcal{T}_{\left[
P_{i}P_{i+1}\right] }$. The module 
\begin{equation}
\left| \mathbf{P}_{i}\mathbf{P}_{i+1}\right| =\sqrt{2\sigma \left(
P_{i},P_{i+1}\right) }=\mu ,\qquad i=0,\pm 1,\pm 2...  \label{e8.5}
\end{equation}
of the vector $\mathbf{P}_{i}\mathbf{P}_{i+1}$ is the geometrical mass $\mu $%
, which is connected with the conventional (physical) mass $m$ by means of
the relation 
\begin{equation}
m=b\mu ,\qquad b\approx 10^{-17}\text{g/cm}  \label{e8.6}
\end{equation}
where $b$ is an universal constant. Analogously, the physical momentum $%
\mathbf{p}_{i}$ is connected with the geometrical momentum $\mathbf{P}_{i}%
\mathbf{P}_{i+1}$ by means of the relation 
\begin{equation}
\mathbf{p}_{k}=b\left( \mathbf{P}_{i}\mathbf{P}_{i+1}\right) _{k}=b\left( 
\mathbf{P}_{i}\mathbf{P}_{i+1}.\mathbf{Q}_{-1}\mathbf{Q}_{k}\right) ,\qquad
k=0,1,2,3  \label{e8.7}
\end{equation}
Here $\left( \mathbf{P}_{i}\mathbf{P}_{i+1}\right) _{k}$ are covariant
coordinates of the vector $\mathbf{P}_{i}\mathbf{P}_{i+1}$ in some
coordinate system with basic vectors $\mathbf{e}_{k}=\mathbf{Q_{-1}Q}_{k}$, $%
k=0,1,2,3$. The coordinate system is determined by five points $\left\{
Q_{-1},Q_{0},Q_{1},Q_{2},Q_{3},\right\} $ with origin at the point $Q_{-1}$.
The scalar product $\left( \mathbf{P}_{i}\mathbf{P}_{i+1}.\mathbf{Q_{-1}Q}%
_{k}\right) $ is defined by the relation 
\begin{equation}
\left( \mathbf{P}_{i}\mathbf{P}_{i+1}.\mathbf{Q_{-1}Q}_{k}\right) =\sigma
\left( P_{i},Q_{k}\right) +\sigma \left( P_{i+1},Q_{-1}\right) -\sigma
\left( P_{i},Q_{-1}\right) -\sigma \left( P_{i+1},Q_{k}\right)  \label{e8.8}
\end{equation}
This definition coincide with conventional definition of the scalar product
in the proper Euclidean space, or in the pseudo-Euclidean one. It is
described in terms of the world function.

If the broken line (\ref{e8.2}) describes the free particle motion, the
momenta of adjacent links are parallel: $\mathbf{P}_{i-1}\mathbf{P}%
_{i}\uparrow \uparrow \mathbf{P}_{i}\mathbf{P}_{i+1}$, $i=0,\pm 1,\pm 2....$
Definition of the parallelism of two vectors $\mathbf{P}_{i-1}\mathbf{P}%
_{i}\ $and $\mathbf{P}_{i}\mathbf{P}_{i+1}$ have the form 
\begin{equation}
\mathbf{P}_{i-1}\mathbf{P}_{i}\uparrow \uparrow \mathbf{P}_{i}\mathbf{P}%
_{i+1}:\qquad \left( \mathbf{P}_{i-1}\mathbf{P}_{i}.\mathbf{P}_{i}\mathbf{P}%
_{i+1}\right) -\left| \mathbf{P}_{i-1}\mathbf{P}_{i}\right| \cdot \left| 
\mathbf{P}_{i}\mathbf{P}_{i+1}\right| =0,\qquad i=0,\pm 1,\pm 2...
\label{e8.9}
\end{equation}

Subtracting relations (\ref{e8.9}) for $i$ and $i+1$ and using the relations
(\ref{e8.5}), (\ref{e8.8}), we obtain from (\ref{e8.9}) 
\begin{equation}
\frac{\left( \mathbf{P}_{i}\mathbf{P}_{i+1}.\mathbf{P}_{i+1}\mathbf{P}%
_{i+2}\right) -\left( \mathbf{P}_{i-1}\mathbf{P}_{i}.\mathbf{P}_{i}\mathbf{P}%
_{i+1}\right) }{\mu ^{2}}=0  \label{e8.10}
\end{equation}
The condition (\ref{e8.10}) is slighter, than the condition (\ref{e8.9}),
but it is interesting in the sense, that lhs of (\ref{e8.10}) describes
''discrete derivative'' of the cosine of the angle between the adjacent
links. According to (\ref{e8.10}) this cosine is constant, besides,
according to (\ref{e8.9}) this cosine is equal to $1$.

Another form of the conditions (\ref{e8.9}) can be obtained, if we use the
definition of the scalar product (\ref{e8.8}) and condition (\ref{e8.5}). We
obtain instead of (\ref{e8.9}) 
\begin{equation}
\left| \mathbf{P}_{i-1}\mathbf{P}_{i+1}\right| =2\mu ,\qquad i=0,\pm 1,\pm
2...  \label{e8.11}
\end{equation}
We can conclude from (\ref{e8.11}) as well as from (\ref{e8.9}), that for
the free particle the broken line (\ref{e8.2}) is a timelike straight line.

The pure geometrical description is useful in the sense that quantum effects
can be taken into account by means of a simple change of the space-time
geometry. We declare that the real space-time geometry is the Minkowski
geometry only approximately. The real space-time geometry is determined by
the world function $\sigma _{\mathrm{d}}$ 
\begin{equation}
\sigma _{\mathrm{d}}=\sigma _{\mathrm{M}}+D\left( \sigma _{\mathrm{M}%
}\right) ,\qquad D\left( \sigma _{\mathrm{M}}\right) =\left\{ 
\begin{array}{ll}
\sigma _{\mathrm{M}}+d & \text{if\ }\sigma _{0}<\sigma _{\mathrm{M}} \\ 
\sigma _{\mathrm{M}} & \text{if\ }\sigma _{\mathrm{M}}<0
\end{array}
\right.  \label{e8.12}
\end{equation}
where $d\geq 0$ and $\sigma _{0}>0$ are some constants. The quantity $\sigma
_{\mathrm{M}}$ is the world function in the Minkowski space-time geometry $%
\mathcal{G}_{\mathrm{M}}$, defined by the relation (\ref{e8.4}). Values of
the function $D\left( \sigma _{\mathrm{M}}\right) $ in interval $\left(
0,\sigma _{0}\right) $ are of no importance, provided geometrical mass $\mu $
of the particle satisfies the condition $\mu \geq \sqrt{2\sigma _{0}}$. The
world function $\sigma _{\mathrm{d}}$ describes geometry $\mathcal{G}_{%
\mathrm{d}}$ of distorted space-time $V_{\mathrm{d}}$, which is
non-Riemannian, if the distortion $d>0$. The geometry $\mathcal{G}_{\mathrm{d%
}}$ is uniform and isotropic, as well as the Minkowski geometry.

The geometry of the distorted space-time $V_{\mathrm{d}}$ is nondegenerate
in the sense, that any link $\mathcal{T}_{\left[ P_{i}P_{i+1}\right] }$,
determined by the relation (\ref{e8.3}) is a 3-dimensional surface (hollow
tube) with the characteristic width $\sqrt{d}$. (See figure 1). If $%
d\rightarrow 0$ the tube degenerates to a segment of the one-dimensional
straight line. Such a situation is connected with the fact that the
definition (\ref{e8.9}) of parallelism is one equation. This equation
determines the set of vectors $\mathbf{P}_{i}\mathbf{P}_{i+1}$, which are
parallel to the fixed vector $\mathbf{P}_{i-1}\mathbf{P}_{i}$, as a set of
points $P_{i+1}$, which satisfy the equation (\ref{e8.9}). In general, this
set is a three-dimensional surface, which degenerates into a one-dimensional
line in the case of the Minkowski geometry and timelike vector $\mathbf{P}%
_{i-1}\mathbf{P}_{i}$. As a result position of the point $P_{i+1}$ in $%
\mathcal{G}_{\mathrm{d}}$ appears to be indefinite, even if the position of
points $P_{i-1}$ and $P_{i}$ is fixed. The shape of world tube $\mathcal{T}_{%
\mathrm{br}}$ appears to be stochastic. The stochasticity intensity depends
on the length $\mu $ of the link. The shorter the length, the larger is
stochasticity, because the characteristic wobble angle is of the order $%
\sqrt{d}/\mu $. If we set 
\begin{equation}
d=\frac{\hbar }{2bc}  \label{e8.14}
\end{equation}
where $\hbar $ is the quantum constant, $c$ is the speed of the light, and $%
b $ is the universal constant, defined by the relation (\ref{e8.6}), the
statistical description of stochastic WLs leads to the quantum description
(in terms of the Schr\"{o}dinger, or Klein-Gordon equation) \cite{R91,R002}.

Note that the statistical description of stochastic WLs leads to a more
general description, than the quantum description. The quantum description
is only the simpler part of the statistical description, which can be
reduced to linear differential equations for the wave function. The
remaining part of the statistical description, which is not reduced to
linear differential equations, has not been investigated, in general.

Thus, the world line of a particle without structure is the geometrical
object, described as a chain (\ref{e8.2}), consisting of similar 1D links $%
\mathcal{T}_{\left[ P_{i}P_{i+1}\right] }$. But, maybe, there exist the
chains (broken tubes), consisting of 2D geometrical objects $\mathcal{T}_{%
\left[ P_{i}P_{i+1}Q_{i}\right] }$, 
\begin{equation}
\mathcal{T}_{\mathrm{br}}=\bigcup\limits_{i}\mathcal{T}_{\left[
P_{i}P_{i+1}Q_{i}\right] }  \label{e8.15}
\end{equation}
where $\mathcal{T}_{\left[ P_{i}P_{i+1}Q_{i}\right] }$ are triangles with
vertices at points $P_{i}$, $P_{i+1}$, $Q_{i}$. This broken tube is shown in
figure 2. There are two versions of the tube (\ref{e8.15}). The version
''dragon'' describes the broken tube (\ref{e8.15}) as consisting of
triangles. The version ''ladder'', containing the same characteristic
points, describes the broken tube (\ref{e8.15}) as consisting of two world
lines $\mathcal{T}_{\mathrm{1}}$ and $\mathcal{T}_{\mathrm{2}}$ 
\begin{equation}
\mathcal{T}_{\mathrm{1}}=\bigcup\limits_{i}\mathcal{T}_{\left[ P_{i}P_{i+1}%
\right] },\qquad \mathcal{T}_{\mathrm{2}}=\bigcup\limits_{i}\mathcal{T}_{%
\left[ Q_{i}Q_{i+1}\right] }  \label{e8.15a}
\end{equation}
Any link $\mathcal{T}_{\left[ P_{i}P_{i+1}\right] }$ of $\mathcal{T}_{%
\mathrm{1}}$ is connected with the link $\mathcal{T}_{\left[ Q_{i}Q_{i+1}%
\right] }$ of $\mathcal{T}_{\mathrm{2}}$ by means of geometrical couplings.
The geometrical coupling acts on $\mathcal{T}_{\mathrm{1}}$ in such a way
that $\mathcal{T}_{\mathrm{1}}$ becomes to be a helix (more exactly all
break points of $\mathcal{T}_{\mathrm{1}}$ lie on the helix (\ref{e6.31})).
In this case we can say, that the broken tube (\ref{e8.15}) describes the
Dirac particle. To obtain this result we may suppose that points $%
P_{i},Q_{i},P_{i+1},Q_{i+1}\ $lie in one 2-dimensional plane and form a
parallelogram. Besides we suppose that all parallelograms $\left\{
P_{i},Q_{i},P_{i+1},Q_{i+1}\right\} $ are similar for $i=0,\pm 1,\pm 2...$

If we suppose that all parallelograms are similar, and dihedral angles
between any pair of adjacent parallelograms are the same, we obtain that all
break points of $\mathcal{T}_{\mathrm{1}}$ lie on a helix. This statement is
valid also for break points of $\mathcal{T}_{\mathrm{2}}$. Choosing
parameters of the broken tube (\ref{e8.15}) in proper way, we can achieve
that the break points of $\mathcal{T}_{\mathrm{1}}$ and $\mathcal{T}_{%
\mathrm{2}}$ lie on the helix (\ref{e6.31}). In this case we may say, that
we obtain a geometrical description of the classical Dirac particle. This
description is analogous to the simpler case, when the broken tube (\ref
{e8.2}) is a geometrical description of the usual particle without an
internal structure.

At the geometrical description it is useless to ask, why world lines $%
\mathcal{T}_{\mathrm{1}}$ and $\mathcal{T}_{\mathrm{2}}$ of constituents
interact at a distance and why there are no bearers of the ''geometrical
interaction''. In general, it is useless to explain geometrical facts by
means of dynamics, because the geometry is more primary and fundamental,
than any dynamics. At the geometrical description any discussion of the
dynamical confinement problems becomes to be useless. We cannot say
definitely, whether the broken tube (\ref{e8.15}) describes a geometrical
coupling of two constituents, describing by (\ref{e8.15a}), or it describes
a chain of more complicated geometrical objects (triangles). See the left
diagram ''dragon'' in figure 2. Two versions ''dragon'' and ''ladder'
distinguish only in their internal geometric couplings, although the
characteristic points $P_{i},Q_{i}$ are the same in both diagrams.

The ''ladder'' is a two-dimensional band in the space-time, which may be
regarded as a world tube of a one-dimensional open-ended rotating string. We
see that the geometrical model of the Dirac particle opens the door for such
notions of the elementary particle theory as string, confinement, quark.

Note that the line segment $\mathcal{T}_{\left[ P_{i}P_{i+1}\right] }$ is
the simplest geometrical object, determined by two points, the triangle $%
\mathcal{T}_{\left[ P_{i}P_{i+1}Q_{i}\right] }$ is the simplest geometrical
object, determined by three points, and the tetrahedron $\mathcal{T}_{\left[
P_{i}P_{i+1}Q_{i}R_{i}\right] }$ is the simplest geometrical object,
described by four points. If the chain of the line segments $\mathcal{T}_{%
\left[ P_{i}P_{i+1}\right] }$ is associated with the spinless particle, the
chain of triangles $\mathcal{T}_{\left[ P_{i}P_{i+1}Q_{i}\right] }$ is
associated with the Dirac particle, we should expect that there exist the
chain of tetrahedrons $\mathcal{T}_{\left[ P_{i}P_{i+1}Q_{i}R_{i}\right] }$.
Such a chain would be associated with the particles, constructed of three
quarks, because such a chain is associated with the composite particle,
consisting of three constituents. Compare diagrams ''dragon'' and ''ladder''
in figure 3.

Now we formulate mathematically constraints on links of the broken tube (\ref
{e8.15}). For simplicity, we consider the case of three-dimensional
space-time. In this case mathematical constraints are more simple and
demonstrative. Note that the helix axis is directed along the vector $%
\mathbf{P}_{i}\mathbf{Q}_{i}$. 
\begin{equation}
\mathbf{P}_{i}\mathbf{Q}_{i}\uparrow \uparrow \mathbf{P}_{i+1}\mathbf{Q}%
_{i+1},\qquad i=0,\pm 1,\pm 2...  \label{e8.16}
\end{equation}
It means that the vector $\mathbf{P}_{i}\mathbf{Q}_{i}\mathbf{\ }$ is to be
timelike 
\begin{equation}
\left| \mathbf{P}_{i}\mathbf{Q}_{i}\right| ^{2}=\left| \mathbf{P}_{i+1}%
\mathbf{Q}_{i+1}\right| ^{2}=\mu ^{2}>0,\qquad i=0,\pm 1,\pm 2...
\label{e8.16a}
\end{equation}
because the helix axis in the case (\ref{e6.31}) is described by the
timelike momentum vector. It follows from relations (\ref{e8.16}), (\ref
{e8.16a}) that the points $P_{i},Q_{i},P_{i+1},Q_{i+1}\ $lie in one
2-dimensional plane and form a parallelogram. Orientation of the
parallelogram in the space-time coincides with the orientation of triangle $%
\mathcal{T}_{\left[ P_{i}Q_{i}P_{i+1}\right] }$. The triangle $\mathcal{T}_{%
\left[ P_{i}Q_{i}P_{i+1}\right] }$ is described by the second order
multivector $\overrightarrow{P_{i}Q_{i}P_{i+1}}\equiv \mathbf{P}_{i}\mathbf{Q%
}_{i}\mathbf{P}_{i+1}$ which is defined as the ordered set $\left\{
P_{i},Q_{i},P_{i+1}\right\} $ of three points \cite{R2002}.

The scalar product of two second order mulivectors $\overrightarrow{\mathcal{%
P}^{2}}\equiv \overrightarrow{P_{0}P_{1}P_{2}}=\left\{
P_{0},P_{1},P_{2}\right\} $ and $\overrightarrow{\mathcal{Q}^{2}}\equiv 
\overrightarrow{Q_{0}Q,Q_{2}}=\left\{ Q_{0},Q_{1},Q_{2}\right\} $ is defined
by the relation \cite{R2002} 
\begin{equation}
\left( \overrightarrow{\mathcal{P}^{2}}.\overrightarrow{\mathcal{Q}^{2}}%
\right) =\det \left\| \left( \overrightarrow{P_{0}P_{i}}.\overrightarrow{%
Q_{0}Q_{k}}\right) \right\| ,\qquad i,k=1,2  \label{e8.18}
\end{equation}
The module $\left| \overrightarrow{\mathcal{P}^{2}}\right| $ of the second
order multivector $\overrightarrow{\mathcal{P}^{2}}$ is defined by the
relation 
\begin{equation}
\left| \overrightarrow{\mathcal{P}^{2}}\right| ^{2}=\left( \overrightarrow{%
\mathcal{P}^{2}}.\overrightarrow{\mathcal{P}^{2}}\right) =\det \left\|
\left( \overrightarrow{P_{0}P_{i}}.\overrightarrow{P_{0}P_{k}}\right)
\right\| ,\qquad i,k=1,2  \label{e8.17a}
\end{equation}

Cosine of the dihedral angle $\theta $ between two second order multivector $%
\overrightarrow{\mathcal{P}^{2}}$ and $\overrightarrow{\mathcal{Q}^{2}}$ is
determined by the relation 
\begin{equation}
\cos \theta =\frac{\left( \overrightarrow{\mathcal{P}^{2}}.\overrightarrow{%
\mathcal{Q}^{2}}\right) }{\left| \overrightarrow{\mathcal{P}^{2}}\right|
\left| \overrightarrow{\mathcal{Q}^{2}}\right| }  \label{e8.17b}
\end{equation}
The dihedral angle between any two adjacent triangles $\mathcal{T}_{\left[
P_{i}Q_{i}P_{i+1}\right] }$ and $\mathcal{T}_{\left[ P_{i+1}Q_{i+1}P_{i+2}%
\right] }$ is the same for all pairs of triangles. As far as all triangles
are equal, we have in addition to (\ref{e8.16a}) the following relations 
\begin{eqnarray}
\left| \mathbf{P}_{i}\mathbf{P}_{i+1}\right| &=&\left| \mathbf{P}_{i+1}%
\mathbf{P}_{i+2}\right| ,\qquad \left( \mathbf{P}_{i}\mathbf{P}_{i+1}.%
\mathbf{P}_{i}\mathbf{Q}_{i}\right) =\left( \mathbf{P}_{i+1}\mathbf{P}_{i+2}.%
\mathbf{P}_{i+1}\mathbf{Q}_{i+1}\right) ,  \label{e8.17c} \\
i &=&0,\pm 1,\pm 2...  \nonumber
\end{eqnarray}
The adjacent triangles $\mathcal{T}_{\left[ P_{i}Q_{i}P_{i+1}\right] }$ and $%
\mathcal{T}_{\left[ P_{i+1}Q_{i+1}P_{i+2}\right] }$ are equal and, hence, 
\begin{equation}
\left| \overrightarrow{P_{i}Q_{i}P_{i+1}}\right| =\left| \overrightarrow{%
P_{i+1}Q_{i+1}P_{i+2}}\right| ,\qquad i=0,\pm 1,\pm 2...  \label{e8.17d}
\end{equation}
Then (\ref{e8.17d}) becomes to be a formal mathematical corollary of
constraints (\ref{e8.16a}) and (\ref{e8.17c}). Taking into account (\ref
{e8.17c}) and (\ref{e8.17b}), the condition of the dihedral angle constancy
is written in the form 
\begin{equation}
\left( \overrightarrow{P_{i}Q_{i}P_{i+1}}.\overrightarrow{%
P_{i+1}Q_{i+1}P_{i+2}}\right) =\left( \overrightarrow{P_{i+1}Q_{i+1}P_{i+2}}.%
\overrightarrow{P_{i+2}Q_{i+2}P_{i+3}}\right) ,\qquad i=0,\pm 1,\pm 2...
\label{e8.17}
\end{equation}

Finally, we must add expression for the link $\mathcal{T}_{\left[
P_{i}Q_{i}P_{i+1}\right] }$, which describes the set of points inside the
triangle with vortices at the points $P_{i},Q_{i},P_{i+1}$. In the case of
arbitrary geometry this expression is rather bulky \cite{R98}. Let $\mathcal{%
P}^{2}$=$\left\{ P_{0},P_{1},P_{2}\right\} $. Then the set of points $R$
inside the triangle with vertices at the points $P_{0},P_{1},P_{2}$ is
described by the relation 
\begin{equation}
\mathcal{T}\left[ \mathcal{P}^{2}\right] \equiv \mathcal{T}_{\left[ \mathcal{%
P}^{2}\right] }=\left\{ R|F_{3}\left( R,\mathcal{P}^{2}\right)
=0\bigwedge_{l=0}^{l=2}S_{l}\geq 0\right\}  \label{e8.19}
\end{equation}
where 
\begin{equation}
F_{3}\left( R,\mathcal{P}^{2}\right) =\det \left| \left| \left( \mathbf{RP}%
_{i}.\mathbf{RP}_{k}\right) \right| \right| ,\qquad i,k=0,1,2  \label{e8.20}
\end{equation}
\begin{eqnarray}
S_{0} &=&\left| \overrightarrow{\mathcal{P}^{2}}\right| ^{-2}\left( 
\overrightarrow{\mathcal{P}^{2}}.\overrightarrow{RP_{1}P_{2}}\right) ,\qquad
S_{1}=\left| \overrightarrow{\mathcal{P}^{2}}\right| ^{-2}\left( 
\overrightarrow{\mathcal{P}^{2}}.\overrightarrow{P_{0}RP_{2}}\right) , 
\nonumber \\
S_{2} &=&\left| \overrightarrow{\mathcal{P}^{2}}\right| ^{-2}\left( 
\overrightarrow{\mathcal{P}^{2}}.\overrightarrow{P_{0}P_{1}R}\right)
\label{e8.21}
\end{eqnarray}
The equation (\ref{e8.20}) describes the two-dimensional plane, determined
by three points $\mathcal{P}^{2}=\left\{ P_{0},P_{1},P_{2}\right\} $.
According the first equation (\ref{e8.21}) the condition $S_{0}\geq 0$ means
that cosine of the dihedral angle between the triangles $\left\{
P_{0},P_{1},P_{2}\right\} $ and $\left\{ R,P_{1},P_{2}\right\} $ is
nonnegative. Hence, the points $R$ and $P_{0}$ are laid in the plane $%
\mathcal{T}_{P_{0}P_{1}P_{2}}$ to one side of the \qquad \qquad \qquad
\qquad \qquad \qquad \qquad \qquad \qquad \qquad \qquad \qquad \qquad \qquad
\qquad \qquad \qquad \qquad \qquad \qquad \qquad \qquad \qquad \qquad \qquad
\qquad \qquad \qquad \qquad \qquad \qquad \qquad \qquad \qquad \qquad \qquad
\qquad \qquad \qquad \qquad \qquad \qquad \qquad \qquad \qquad \qquad \qquad
\qquad \qquad \qquad \qquad \qquad \qquad \qquad \qquad \qquad \qquad \qquad
straight $\mathcal{T}_{P_{1}P_{2}}$. In a like manner the condition $%
S_{1}\geq 0$ means that the points $R$ and $P_{1}$ are laid in the plane $%
\mathcal{T}_{P_{0}P_{1}P_{2}}$ to one side inside of the straight $\mathcal{T%
}_{P_{0}P_{2}}$.

Let us determine connection between the parameters $\gamma ,m$ of world line
(\ref{e6.31}) of the Dirac particle and the parameters of triangles $%
\mathcal{T}_{\left[ P_{i}P_{i+1}Q_{i}\right] }$, constituting the world tube
(\ref{e8.15}). Projection of the tube (\ref{e8.15}) onto the 2-plane
orthogonal to the parallel vectors $\mathbf{P}_{i}\mathbf{Q}_{i}$, $i=0,\pm
1,\pm 2...$ is shown in figure 4 (remember that we are considering now the
3D-case). Points $P_{i},Q_{i}$ are projected into one point. The angle $%
\theta $ is the dihedral angle between triangles $P_{i}P_{i+1}Q_{i}$ and $%
P_{i+1}P_{i+2}Q_{i+1}$. The point $O$ is the center of the circle, where the
points $P_{i}$, $Q_{i}$, $i=0,\pm 1,\pm 2...$ are placed. The radius $%
R=\left| \mathbf{OP}_{i}\right| $ of this circle is identified with the
radius 
\begin{equation}
a=\frac{\hbar \gamma \sqrt{\gamma ^{2}-1}}{2m}  \label{e8.22}
\end{equation}
of the world line (\ref{e6.31}). It follows from the figure 4, that 
\begin{equation}
R=\frac{2\left| \mathbf{P}_{0}\mathbf{P}_{1}\right| }{\sin \frac{\Delta
\varphi }{2}}=\frac{2\left| \mathbf{P}_{0}\mathbf{P}_{1}\right| }{\cos \frac{%
\theta }{2}}  \label{e8.23}
\end{equation}

Displacement $\Delta t$ of the point $P_{i}$ along the helix axis
corresponds to the angle $\Delta \varphi =\pi -\theta $. It is determined by
the relation 
\begin{equation}
\Delta t=\frac{\left( \mathbf{P}_{1}\mathbf{P}_{2}.\mathbf{P}_{1}\mathbf{Q}%
_{1}\right) }{\left| \mathbf{P}_{1}\mathbf{Q}_{1}\right| }  \label{e8.24}
\end{equation}
The the angular frequency is determined by the relation 
\begin{equation}
\omega =\frac{\Delta \varphi }{\Delta t}=\frac{\left( \pi -\theta \right)
\left| \mathbf{P}_{1}\mathbf{Q}_{1}\right| }{\left( \mathbf{P}_{1}\mathbf{P}%
_{2}.\mathbf{P}_{1}\mathbf{Q}_{1}\right) }  \label{e8.25}
\end{equation}
The angular frequency (\ref{e8.25}) should be identified with the angular
frequency $\Omega $, defined by the relation (\ref{e6.29}). It gives 
\begin{equation}
\frac{2m}{\hbar \gamma ^{2}}=\frac{\left( \pi -\theta \right) \left| \mathbf{%
P}_{1}\mathbf{Q}_{1}\right| }{\left( \mathbf{P}_{1}\mathbf{P}_{2}.\mathbf{P}%
_{1}\mathbf{Q}_{1}\right) },  \label{e8.26}
\end{equation}
or 
\begin{equation}
\gamma =\sqrt{\frac{2m\left( \mathbf{P}_{1}\mathbf{P}_{2}.\mathbf{P}_{1}%
\mathbf{Q}_{1}\right) }{\hbar \left( \pi -\theta \right) \left| \mathbf{P}%
_{1}\mathbf{Q}_{1}\right| }}  \label{e8.27}
\end{equation}
where $\gamma $, as well as $\theta $ are parameters, defining internal
motion of the Dirac particle. The parameter $\theta $ is determined by the
mutual disposition of links (triangles) in the chain (\ref{e8.15}).

In the 4D space-time constraints on the broken tube (\ref{e8.15}) are to be
chosen in such a way that we obtain the diagram of figure 4 for projections
of points $P_{i},Q_{i}$ on some two-dimensional plane which is orthogonal to
vectors $\mathbf{P}_{i}\mathbf{Q}_{i}$ and some spacelike vector $\mathbf{P}%
_{i}\mathbf{S}_{i}$. All vectors $\mathbf{P}_{i}\mathbf{S}_{i}$ are similar,
and we consider them as one vector $\mathbf{\xi }$. The vector $\mathbf{P}%
_{i}\mathbf{S}_{i}$ satisfies the relations 
\begin{equation}
\mathbf{P}_{i}\mathbf{S}_{i}=\mathbf{P}_{i+1}\mathbf{S}_{i+1},\qquad \left( 
\mathbf{P}_{i}\mathbf{S}_{i}.\mathbf{P}_{i}\mathbf{Q}_{i}\right) =0,\qquad
\left| \mathbf{P}_{i}\mathbf{S}_{i}\right| ^{2}=-1,\qquad i=0,\pm 1,\pm 2...
\label{e8.28}
\end{equation}
In the coordinate system, where 
\begin{equation}
\mathbf{P}_{i}\mathbf{Q}_{i}=\left\{ \mu ,0,0,0\right\} ,\qquad \mathbf{P}%
_{i}\mathbf{S}_{i}=\mathbf{\xi }=\left\{ 0,0,0,1\right\}  \label{e8.29}
\end{equation}
vectors $\mathbf{P}_{i}\mathbf{P}_{i+1}$, $\mathbf{Q}_{i}\mathbf{Q}_{i+1}$
have components 
\begin{equation}
\mathbf{P}_{i}\mathbf{P}_{i+1}=\mathbf{Q}_{i}\mathbf{Q}_{i+1}=\left\{
0,u_{1},u_{2}.0\right\}  \label{e8.30}
\end{equation}

Thus, relations (\ref{e8.16}), (\ref{e8.16a}), (\ref{e8.17c}) and (\ref
{e8.17}) are conditions of the geometrical description of the classical
Dirac particle. We can return from them to dynamic description. If we
replace the Minkowski world function in relations (\ref{e8.16}), (\ref
{e8.16a}), (\ref{e8.17c}) and (\ref{e8.17}) by the world function (\ref
{e8.12}), (\ref{e8.14}), we obtain stochastic world tubes (\ref{e8.15}).
Introducing statistical description of the stochastic world tubes, we obtain
some version of the quantum description. Does this description coincide with
the description in terms of the Dirac equation? Maybe, but it is not
necessarily. There are some reasons for such a hesitation. In the case of
the spinless particle, described by the world tube (\ref{e8.2}), the
statistical description leads to a more general description, than the
conventional quantum description in terms of the Schr\"{o}dinger equation.
The quantum description is only a special part of the general statistical
description. Second, we cannot be sure, that helices (\ref{e6.31}) can be
obtained only at the conditions (\ref{e8.16}), (\ref{e8.17c}) and (\ref
{e8.17}). Maybe, there are another conditions, which lead to the helices (%
\ref{e6.31}).

Nevertheless, the approach, founded on the a choice of the proper space-time
geometry and proper geometrical objects as candidates for descriptions of
elementary particles seems to be rather promising. In this case we do not
use enigmatic quantum principles, we do not invent exotic properties of
particles and of space-time. We do not invent new hypotheses, we simply look
for the proper space-time geometry, and the proper geometrical objects in
the set of known geometries and in the set of geometrical objects with known
properties. Unfortunately, we cannot restrict ourselves by consideration of
the space-time geometry, because now we are able to construct a statistical
description only in the framework of dynamics (but not in the framework of
geometry). Nevertheless, such a dynamical problem as confinement does not
arise, if we start from geometry. Besides, it seems rather probable, that
masses of elementary particles are determined mainly by their geometric
structure, whereas contribution of quantum effects, conditioned by a
distortion  (\ref{e8.12}) of the space-time geometry is negligible.

The long-range action is another problem of the relativistic rotator. This
is also a dynamical problem, which is absent in the geometrical model.

The relations (\ref{e8.16a}), and conditions (\ref{e8.17c}), describing that
all links (triangles) must be similar seem to be rather reasonable.

In this paper we have investigated well-known dynamic system $\mathcal{S}_{%
\mathrm{D}}$. We used methods of the model conception of quantum phenomena
(MCQP) \cite{R004}. We did not used any additional suppositions.
Furthermore, we have removed all quantum principles and have not use them in
our investigations. Results of investigation of the well-known dynamic
system $\mathcal{S}_{\mathrm{D}}$ appeared to be unexpected and encouraging.
We have came to the approach, containing a series of notions of the
contemporary elementary particles such as string, quark, confinement.
Appearance of these concepts is not connected with any additional
hypotheses. It is rather reasonable, because their appearance is connected
with such a fundamental structure as the space-time geometry.

Why have we obtain these results, which could not be obtained on the basis
of the quantum principles? The answer is rather unexpected. Conventional
theory of physical phenomena in microcosm contains mistakes, which are
compensated by means of the quantum principles. MCQP these mistakes are
corrected, and there is no necessity to compensate them. The quantum
principles became to be unnecessary. As a result the theory of the microcosm
phenomena and its methods become simple and reasonable.

\newpage \renewcommand{\theequation}{\Alph{section}.\arabic{equation}} %
\renewcommand{\thesection}{\Alph{section}} \setcounter{section}{0} %
\centerline{\Large \bf Mathematical Appendices}

\section{Calculation of Lagrangian}

Let us calculate the expression 
\begin{equation}
{\frac{i}{2}}\hbar \bar{\psi}\gamma ^{l}\partial _{l}\psi +\text{h.c}%
=F_{1}+F_{2}+F_{3}+F_{4}  \label{c8.1}
\end{equation}
where the following designations are used 
\begin{equation}
F_{1}={\frac{i}{2}}\hbar \psi ^{\ast }\left( \left( \partial _{0}-i\gamma
_{5}\mathbf{\sigma \nabla }\right) i\varphi \right) \psi +\text{h.c.}
\label{c8.2}
\end{equation}
\begin{equation}
F_{2}={\frac{i}{2}}\hbar \psi ^{\ast }\left( \left( \partial _{0}-i\gamma
_{5}\mathbf{\sigma \nabla }\right) \left( \frac{1}{2}\gamma _{5}\kappa
\right) \right) \psi +\text{h.c.}  \label{c8.3}
\end{equation}
\begin{eqnarray}
F_{3} &=&\frac{i}{2}\hbar A^{2}\Pi \left( \left( \mathbf{\sigma n}\right)
\exp \left( -i\gamma _{5}\mathbf{\sigma \eta }\right) \left( \mathbf{\sigma n%
}\right) \right) \exp \left( -\frac{i\pi }{2}\mathbf{\sigma n}\right) 
\nonumber \\
&&\times (\partial _{0}-i\gamma _{5}\mathbf{\sigma \nabla })\exp \left( 
\frac{i\pi }{2}\mathbf{\sigma n}\right) \Pi +\text{h.c.}  \label{c8.4}
\end{eqnarray}
\begin{equation}
F_{4}=\frac{i}{2}\hbar A^{2}\Pi \exp \left( -\frac{i}{2}\gamma _{5}\mathbf{%
\Sigma }\mathbf{\eta }\right) (\partial _{0}-i\gamma _{5}\mathbf{\Sigma
\nabla })\left( -\frac{i}{2}\gamma _{5}\mathbf{\Sigma }\mathbf{\eta }\right)
\Pi +\text{h.c.}  \label{c8.5}
\end{equation}
In the last relation the matrix $\mathbf{\Sigma }$ is not differentiated.

Using definitions of $j^{l}$ and $S^{l}$, the expression $F_{1}$ and $F_{2}$
reduce to the form 
\begin{equation}
F_{1}={\frac{i}{2}}\hbar \psi ^{\ast }\left( \left( \partial _{0}-i\gamma
_{5}\mathbf{\sigma \nabla }\right) i\varphi \right) \psi +\text{h.c.}%
=-j^{l}\partial _{l}\varphi \Pi  \label{b5.5}
\end{equation}
\begin{eqnarray*}
F_{2} &=&{\frac{i}{2}}\hbar \psi ^{\ast }\left( \left( \partial _{0}-i\gamma
_{5}\mathbf{\sigma \nabla }\right) \left( \frac{1}{2}\gamma _{5}\kappa
\right) \right) \psi +\text{h.c} \\
&=&{\frac{i}{2}}\hbar \psi ^{\ast }\gamma _{5}\gamma ^{l}\partial _{l}\left( 
\frac{1}{2}\gamma _{5}\kappa \right) \psi +\text{h.c}
\end{eqnarray*}
\begin{equation}
F_{2}=-{\frac{1}{2}}\hbar S^{l}\partial _{l}\kappa \Pi  \label{b5.6}
\end{equation}

\begin{eqnarray*}
F_{3} &=&\frac{i}{2}\hbar A^{2}\Pi \exp \left( -\frac{i\pi }{2}\mathbf{%
\sigma n}\right) \exp \left( -i\gamma _{5}\mathbf{\sigma \eta }\right)
(\partial _{0}-i\gamma _{5}\mathbf{\sigma }\nabla )\exp \left( \frac{i\pi }{2%
}\mathbf{\sigma n}\right) \Pi +\text{h.c.} \\
&=&\frac{i}{2}\hbar j^{l}\Pi \sigma _{\alpha }\sigma _{\beta }n^{\alpha
}\partial _{l}n^{\beta }\Pi +\text{h.c.}=\frac{i}{2}\hbar j^{l}n^{\alpha
}\partial _{l}n^{\beta }\Pi \left( \delta _{\alpha \beta }+i\varepsilon
_{\alpha \beta \gamma }\sigma _{\gamma }\right) \Pi +\text{h.c.} \\
&=&\frac{i}{2}\hbar j^{l}\left( n^{\alpha }\partial _{l}n^{\alpha
}+i\varepsilon _{\alpha \beta \gamma }n^{\alpha }\partial _{l}n^{\beta
}z^{\gamma }\right) \Pi +\text{h.c.}
\end{eqnarray*}
As far as $\mathbf{n}^{2}=1$, one obtains 
\begin{equation}
n^{\alpha }\partial _{l}n^{\alpha }=0  \label{b5.8}
\end{equation}
Besides it follows from (\ref{a3.21}) that 
\begin{equation}
\mathbf{n}=\frac{\mathbf{\sigma }+\mathbf{z}}{\sqrt{2\left( 1+\mathbf{\sigma
z}\right) }}  \label{b5.9}
\end{equation}

Then 
\begin{equation}
F_{3}=-\hbar j^{l}\left( \varepsilon _{\alpha \beta \gamma }n^{\alpha
}\partial _{l}n^{\beta }z^{\gamma }\right) \Pi =-\frac{\hbar j^{l}}{2\left(
1+\mathbf{\xi z}\right) }\varepsilon _{\alpha \beta \gamma }\xi ^{\alpha
}\partial _{l}\xi ^{\beta }z^{\gamma }\Pi  \label{c8.9}
\end{equation}
Calculation of $F_{4}$ leads to the following result 
\begin{eqnarray*}
F_{4} &=&\frac{i}{2}\hbar A^{2}\Pi \exp \left( -\frac{i}{2}\gamma _{5}%
\mathbf{\Sigma }\mathbf{\eta }\right) (\partial _{0}-i\gamma _{5}\mathbf{%
\Sigma \nabla })\exp \left( -\frac{i}{2}\gamma _{5}\mathbf{\Sigma }\mathbf{%
\eta }\right) \Pi +\text{h.c.} \\
&=&\frac{i}{2}\hbar A^{2}\Pi \left( \cosh \frac{\eta }{2}-i\gamma
_{5}v^{\alpha }\Sigma _{\alpha }\sinh \frac{\eta }{2}\right) \\
&&\times (\partial _{0}-i\gamma _{5}\mathbf{\Sigma \nabla })\left( \cosh 
\frac{\eta }{2}-i\gamma _{5}v^{\alpha }\Sigma _{\alpha }\sinh \frac{\eta }{2}%
\right) \Pi +\text{h.c.} \\
&=&\frac{i}{2}\hbar A^{2}\Pi \left( \cosh \frac{\eta }{2}\sinh \frac{\eta }{2%
}\partial _{0}\eta +\sinh ^{2}\frac{\eta }{2}v^{\alpha }\partial
_{0}v^{\beta }\Sigma _{\alpha }\Sigma _{\beta }\right) \Pi +\text{h.c.} \\
&&+\frac{i}{2}\hbar A^{2}\Pi \left( \cosh ^{2}\frac{\eta }{2}\Sigma _{\alpha
}\Sigma _{\beta }v^{\beta }+\sinh ^{2}\frac{\eta }{2}\Sigma _{\beta }\Sigma
_{\alpha }v^{\beta }\right) \partial _{\alpha }\frac{\eta }{2}\Pi +\text{h.c.%
} \\
&&+\frac{i}{2}\hbar A^{2}\Pi \cosh \frac{\eta }{2}\sinh \frac{\eta }{2}%
\Sigma _{\alpha }\Sigma _{\beta }\partial _{\alpha }v^{\beta }\Pi +\text{h.c.%
}
\end{eqnarray*}
\begin{eqnarray*}
F_{4} &=&\frac{i}{2}\hbar A^{2}\Pi \left( \frac{1}{2}\sinh \eta \partial
_{0}\eta +\sinh ^{2}\frac{\eta }{2}v^{\alpha }\partial _{0}v^{\beta
}i\varepsilon _{\alpha \beta \gamma }\Sigma _{\gamma }\right) \Pi +\text{h.c.%
} \\
&&+\frac{i}{2}\hbar A^{2}\Pi \left( \cosh \eta v^{\alpha }+i\varepsilon
_{\beta \alpha \gamma }v^{\beta }\Sigma _{\gamma }\right) \partial _{\alpha }%
\frac{\eta }{2}\Pi +\text{h.c.} \\
&&+\frac{i}{4}\hbar A^{2}\Pi \sinh \eta \left( \partial _{\alpha }v^{\alpha
}+i\varepsilon _{\alpha \beta \gamma }\partial _{\alpha }v^{\beta }\Sigma
_{\gamma }\right) \Pi +\text{h.c.}
\end{eqnarray*}
\begin{eqnarray*}
F_{4} &=&-\hbar A^{2}\left( \sinh ^{2}\frac{\eta }{2}v^{\alpha }\partial
_{0}v^{\beta }\varepsilon _{\alpha \beta \gamma }\xi ^{\gamma }\right) \Pi \\
&&-\frac{1}{2}\hbar A^{2}\Pi \left( \varepsilon _{\beta \alpha \gamma
}v^{\beta }\xi ^{\gamma }\partial _{\alpha }\eta +\sinh \eta \varepsilon
_{\alpha \beta \gamma }\partial _{\alpha }v^{\beta }\xi ^{\gamma }\right) \Pi
\end{eqnarray*}
\begin{equation}
F_{4}=-{\frac{1}{2}}\hbar A^{2}\varepsilon _{\alpha \beta \gamma }\left(
\partial _{\alpha }\eta v^{\beta }+\sinh \eta \partial _{\alpha }v^{\beta
}+2\sinh ^{2}(\frac{\eta }{2})v^{\alpha }\partial _{0}v^{\beta }\right) \xi
^{\gamma }\Pi  \label{c8.10}
\end{equation}

According to the relation (\ref{a3.17}) $A=(j^{l}j_{l})^{1/4}\equiv \rho
^{1/2}$, and relation (\ref{c8.10}) may be written in the form 
\begin{equation}
F_{4}=\left( \frac{\hbar (\rho +j_{0})}{2}\varepsilon _{\alpha \beta \gamma
}\partial ^{\alpha }\frac{j^{\beta }}{(j^{0}+\rho )}-\frac{\hbar }{2(\rho
+j_{0})}\varepsilon _{\alpha \beta \gamma }\left( \partial ^{0}j^{\beta
}\right) j^{\alpha }\right) \xi ^{\gamma }\Pi  \label{c9.1}
\end{equation}

To prove this statement, we substitute the expression of $j^{i}$ via
variables $\mathbf{v,}$ $\eta $ 
\begin{equation}
j^{0}=\rho \cosh \eta ,\qquad j^{\alpha }=\rho \sinh \eta v^{\alpha }
\label{c9.2}
\end{equation}
in (\ref{c9.1}). We obtain 
\begin{eqnarray*}
F_{4} &=&\frac{\hbar \rho (1+\cosh \eta )}{2}\varepsilon _{\alpha \beta
\gamma }\partial ^{\alpha }\frac{\sinh \eta v^{\beta }}{(1+\cosh \eta )}\xi
^{\gamma }\Pi \\
&&-\frac{\hbar \rho }{2(1+\cosh \eta )}\varepsilon _{\alpha \beta \gamma
}\partial ^{0}\left( \sinh \eta v^{\beta }\right) \sinh \eta v^{\alpha }\xi
^{\gamma }\Pi \\
&=&\hbar \rho (\cosh ^{2}\frac{\eta }{2})\varepsilon _{\alpha \beta \gamma
}\partial ^{\alpha }\left( \tanh \frac{\eta }{2}v^{\beta }\right) \xi
^{\gamma }\Pi -\frac{\hbar \rho \sinh ^{2}\eta }{2(1+\cosh \eta )}%
\varepsilon _{\alpha \beta \gamma }\left( \partial ^{0}v^{\beta }\right)
v^{\alpha }\xi ^{\gamma }\Pi \\
&=&\hbar \rho (\cosh ^{2}\frac{\eta }{2})\varepsilon _{\alpha \beta \gamma
}\partial ^{\alpha }\left( \tanh \frac{\eta }{2}v^{\beta }\right) \xi
^{\gamma }\Pi -\hbar \rho \sinh ^{2}\frac{\eta }{2}\varepsilon _{\alpha
\beta \gamma }\left( \partial ^{0}v^{\beta }\right) v^{\alpha }\xi ^{\gamma
}\Pi
\end{eqnarray*}
\[
F_{4}=\frac{\hbar \rho }{2}\varepsilon _{\alpha \beta \gamma }\left(
\partial ^{\alpha }\eta v^{\beta }+\sinh \eta \partial ^{\alpha }v^{\beta
}-2\sinh ^{2}\frac{\eta }{2}\partial ^{0}\left( v^{\beta }\right) v^{\alpha
}\right) \xi ^{\gamma }\Pi 
\]

The obtained relation coincides with the expression (\ref{c8.10}) for $F_{4}$%
, if we take into account that $\partial ^{\alpha }=-\partial _{\alpha }$.

\section{Transformation of equation for variable $\mathbf{\protect\xi }$}

Multiplying equation (\ref{b6.5}) by $(1+\mathbf{z\xi })$ and keeping in
mind $\mathbf{\xi }^{2}=1$ and $\mathbf{z}^{2}=1$, we obtain 
\begin{equation}
\mathbf{\xi }\times \left( -\dot{\mathbf{\xi }}\times \mathbf{z}+{\frac{(%
\mathbf{z}\dot{\mathbf{\xi }})}{2(1+\mathbf{z\ \xi })}}\mathbf{\xi }\times 
\mathbf{z-}\frac{\mathbf{\dot{\xi}}(\mathbf{\xi }\times \mathbf{z})}{2(1+%
\mathbf{z\ \xi })}\mathbf{z}-{\frac{(1+\mathbf{z\ \xi })}{2}}\mathbf{b}%
\right) =0,\qquad \mathbf{b}=-(\dot{\mathbf{x}}\times \ddot{\mathbf{x}})Q
\label{f9.1}
\end{equation}
Two middle terms could be represented as the double vector product 
\begin{equation}
\mathbf{\xi }\times \left( -\dot{\mathbf{\xi }}\times \mathbf{z}+{\frac{1}{%
2(1+\mathbf{z\ \xi })}}\left( \mathbf{\dot{\xi}}\times \left( {(\mathbf{\xi }%
\times \mathbf{z})\times }\mathbf{z}\right) \right) -{\frac{(1+\mathbf{z\
\xi })}{2}}\mathbf{b}\right) =0  \label{f9.2}
\end{equation}
This equation can be rewritten in the form 
\begin{equation}
\mathbf{\xi }\times \left( \dot{\mathbf{\xi }}\times \left( -\mathbf{z+}{%
\frac{\left( \mathbf{z\xi }\right) \mathbf{z-\xi }}{2(1+\mathbf{z\ \xi })}}%
\right) -{\frac{(1+\mathbf{z\ \xi })}{2}}\mathbf{b}\right) =0  \label{f9.3}
\end{equation}
Now calculating double vector products and taking into account that $\mathbf{%
\xi \dot{\xi}}=0$, one obtains 
\begin{equation}
-\dot{\mathbf{\xi }}-\left( \mathbf{\xi }\times \mathbf{b}\right) =0
\label{f9.5}
\end{equation}
or 
\begin{equation}
\dot{\mathbf{\xi }}=\frac{1}{2}\left( \mathbf{\xi }\times (\dot{\mathbf{x}}%
\times \ddot{\mathbf{x}})\right) Q  \label{f9.6}
\end{equation}

\end{document}